\documentclass[]{interact}
\usepackage{lipsum}  
\usepackage{epstopdf}
\usepackage[caption=false]{subfig}

\usepackage[natbibapa,nodoi]{apacite}
\usepackage{graphicx}
\usepackage{float}
\graphicspath{{png/} }
\usepackage{comment}
\usepackage{algorithm}
\usepackage{algorithmic}
\usepackage{array}
\usepackage{multirow}

\newcolumntype{P}[1]{>{\centering\arraybackslash}p{#1}}
\theoremstyle{plain}

\theoremstyle{definition}

\theoremstyle{remark}

\usepackage{xcolor}

\usepackage{graphics}
\usepackage{color}
\definecolor{rev1}{rgb}{0,0,0}

\begin{document}


\title{Nonintrusive reduced order modeling of convective Boussinesq flows}


\author{
\name{Pedram H.~Dabaghian, Shady E.~Ahmed, and Omer San \thanks{CONTACT O. San. Email: osan@okstate.edu}}
\affil{School of Mechanical \& Aerospace Engineering, Oklahoma State University, Stillwater, OK 74078, USA.}
}

\maketitle

\begin{abstract}
In this paper, we formulate three nonintrusive methods and systematically explore their performance in terms of the ability to reconstruct the quantities of interest and their predictive capabilities. 
The methods include deterministic dynamic mode decomposition (DMD), randomized DMD and nonlinear proper orthogonal decomposition (NLPOD). We apply these methods to a convection dominated fluid flow problem governed by the Boussinesq equations. 
\textcolor{rev1}{We analyze the reconstruction results primarily at two different times for considering different noise levels synthetically added into the data snapshots.}
Overall, our results indicate that, with a proper selection of the number of retained modes and neural network architectures, all three approaches make predictions that are in a good agreement with the full order model solution. \textcolor{rev1}{However, we find that the NLPOD approach seems more robust for higher noise levels compared to both DMD approaches.} 
\end{abstract}


\section{Introduction}
Thanks to the advent of modern data acquisition devices such as sensors and their auxiliary transferring, preprocessing, and storing tools, nowadays, the bottleneck is shifting away from the amount of available information to identify complicated physical phenomena. Instead, the challenging part is the required computational effort to find meaningful correlations among different features of the targeted data set. In response to the concern of computing, artificial intelligence (AI) is now being introduced as a futuristic approach that can solve fundamental problems in scientific societies in a variety of fields \citep{vinuesa2020role}.
Moreover, with the popularity of digital twins as a tool that provides online monitoring and flexibility in operation, the need to speed up numerical calculations is felt more \citep{rasheed2020digital,kapteyn2021probabilistic,san2021hybrid}. Despite the significant developments in the computational fluid dynamics (CFD) studies over the last few decades, numerical techniques are still too sluggish to simulate practical flow problems \citep{ahmed2021closures}.

In this regard, there have always been ideas to shrink the data so that the most expressive parts of the underlying system remain unaffected. Recently, reduced order modeling (ROM) has become a promising approach to answer this need as they have successfully delivered reliable results with acceptable accuracy \citep{quarteroni2015reduced,hesthaven2016certified,bui2008model}. By defining quantitative criteria, a successful ROM can differentiate between valuable and redundant information while saving up to two or three orders of magnitude in computing costs \citep{yu2019non}. In ROM, the prevailing assumption at all stages is that the information contained in the data can be presented on a compact solution manifold. First, it might be possible to segregate some meaningful and interpretive information from the less important flow features in such a way that in the first place acceptable accuracy is achieved. Second, there might exist a possibility of retrieving full order data using the retained information. 

Generally, ROMs can be classified based on the level of access to the full order model (FOM) that is required to approximate the system's dynamics. From this point of view, there are intrusive and nonintrusive classes. In the intrusive version, the discretized partial differential equation (PDE) operators are required for the numerical execution of the simplified models \citep{san2015principal,san2015stabilized,yildiz2021intrusive}. In particular, the high-fidelity system is projected onto a reduced basis, which is computed in a variety of ways. One of the most common methodologies to define effective low rank approximations for this purpose is the proper orthogonal decomposition (POD). In POD, the system can be described using a linear subspace which creates a spatial orthogonal basis. Overall, the POD analysis identifies the best set of spatial modes (in the linear sense) for extracting as much information from the flow field data as possible over time. Besides, POD itself is a completely data-based best-linear-fit approach that does not demand prior knowledge of the underlying dynamics. It requires flow field data derived from either numerical simulations or experimental measurements \citep{taira2020modal}. 

After defining the orthogonal bases, a model for their associated coefficients can be built by projecting the FOM operators onto the reduced basis through a process often denoted as Galerkin projection. Despite being intrusive and having high level of accuracy, this method might not be a proper choice for a general nonlinear problem. On one hand, Galerkin-based ROM can yield inaccurate and unstable results when they operate in the under-resolved regime, which is often the case for computational savings purposes \citep{grimberg2020stability,ahmed2020breaking,ahmed2021closures}. On the other hand, they suffer from the huge computational costs that should be allocated to the projection phase as an intrusive ROM method. This issue is even more pronounced when the underlying FOM solvers are out-of-reach or protected by copyrights. There have been many works to tackle this problem by bypassing the projection step \citep{pawar2019deep, aversano2019application, rahman2019nonintrusive, yu2019non, dutta2021pynirom, kramer1991nonlinear, monahan2000nonlinear, hsieh2007nonlinear, otto2019linearly, iwata2020neural, pan2020physics, puligilla2018deep, carlberg2019recovering, phillips2021autoencoder}. There are also other reduction strategies that have been developed to address the drawbacks of intrusive methods such as the empirical interpolation method (EIM) \citep{barrault2004empirical, grepl2007efficient}, the discrete empirical interpolation method (DEIM) \citep{chaturantabut2010nonlinear, cstefuanescu2013pod, xiao2014non}, the trajectory piece-wise linear (TPWL) method \citep{rewienski2006model}, the function sampling method \citep{astrid2008missing, carlberg2011efficient, carlberg2013gnat} and machine learning (ML) \citep{brunton2020machine}. Among the mentioned strategies, ML has been able to attract the attention of many scientists by using various tools such as the long short-term memory (LSTM) neural network, autoencoder (AE), and convolutional autoencoder (CAE) \citep{vlachas2022multiscale, novati2021automating}.

On the other hand, nonintrusive ROM (NIROM) methods are designed to model the effective system's dynamics and reconstruct the high dimensional data set without requiring any access to the full set of governing equations or CFD source codes \citep{pawar2019deep, hampton2018practical, chen2018greedy, xiao2019domain, wang2019non, peherstorfer2016data, xiao2015non, xiao2015non2, xiao2015non3, xiao2016non, xiao2017non, xiao2017towards, heaney2022ai}. \textcolor{rev1}{In other words, nonintrusive
approaches are characterized by the fact that they only require access to the state vectors of the
dynamical system but not to the operators of the system themselves.}  Therefore, NIROM can potentially provide a computationally efficient substitute to its intrusive peer and this well justifies the increasing use of these types of approaches.

One of the ML approaches that has recently been applied on fluid mechanics problems is nonlinear proper orthogonal decomposition (NLPOD) \citep{ahmed2021nonlinear}. In this approach, a set of best linear-fit basis functions are constructed using POD, followed by a combination of feed-forward autoencoder and a time series prediction tool (e.g., LSTM) to evolve the associated modal coefficients. In fact, several ML tools can be applied to detour projecting onto the PDEs. More details on numerous ML approaches are available in \citet{beck2021perspective,vlachas2022multiscale}. In our benchmarking, we closely follow the strategy presented by \citet{ahmed2021nonlinear}.

As an alternative nonintrusive approach, dynamic mode decomposition (DMD) has become a popular model reduction approach in fluid mechanics community due to its capability to deal with dynamical problems wherein the dominant modes can properly interpret a measurable quantity. DMD can be viewed as a data-driven approximation of the Koopman operator, where a linear-based fit is defined using a set of system's observables that encompass the underlying dynamics.
Unlike POD where modes are computed solely based on a reconstruction error minimization criterion, DMD provides a set of modes that grow or decay exponentially in time and oscillate with distinct frequencies. By computing the DMD modes and their associated initial amplitudes and eigenvalues, the reduced order model can estimate the system's behavior at a given time. The DMD technique has been used to address problems with various physics and the reconstructed fields were compared to high fidelity data to measure the strategy's effectiveness \citep{jang2021oscillatory}. As a nonintrusive ROM technique, DMD identifies some basis functions from the full order data through a truncation step for later use in the reconstruction of the system. Since any elimination of the data is accompanied by some error, the truncation phase imposes some limitations on the behavior of DMD. In this case, selecting updatable system parameters can compensate for the truncation-induced errors. 

To decrease truncation errors in dynamical models, \citet{wilson2022data} presented an adaptation of the DMD with control (DMDc) approach that implements adaptive parameters. As an extension to the sparsity promoting DMD introduced by \citet{jovanovic2014sparsity}, the recent work by \citet{pan2021sparsity} focused on mode selection by employing a new multi-task learning method. While DMD is a reliable data-driven method for discovering and interpreting the oscillating modes of a large number of PDEs, it has the aforementioned drawbacks and needs to be improved \citep{rot2022dynamic}. For example, \cite{katrutsa2022extension} showed that the typical DMD might fail to interpret the dynamical behavior of a system with incomplete measurement settings. Despite the fact that decreasing the system's order appears to be crucial for a modal analysis procedure, a study conducted by \citet{li2022parametric} discussed that truncating some low-energy states leads to the omission of some aspects that are important in the system's temporal structure. This underscores the importance of an effective mode selecting method that takes into account a unique spatiotemporal aspect of the system.

In this study, we explore the capability of two different nonintrusive ROM approaches, NLPOD and DMD, in reconstructing the information considering a convective flow problem, which is often regarded as a challenging task for the evaluation of ROM methods. More specifically, we investigate two different variants of DMD methods, namely the deterministic DMD (DDMD) and randomised DMD (RDMD). We note that DMD (being a frequency-based approach) and NLPOD (using an energy-based decomposition) also exploit different mode selecting procedures and consequently yield different reconstruction results. The approaches are evaluated by calculating the root mean squared error (RMSE) of the ROM's predictions with different number of modes, targeting expressing the role of the number of utilized modes in the accuracy of the reconstruction.

The remainder of the paper is organized as follows. In section~\ref{sec:gov}, the two dimensional Boussinesq equations are discussed briefly. The mathematical formulation of the deterministic and randomized versions of DMD is provided in sections~\ref{sec:ddmd} and \ref{sec:rdmd}, respectively. Moreover, the NLPOD method is covered in section~\ref{sec:nlpod}. Results are discussed and analyzed in section~\ref{sec:res}, while concluding remarks are drawn in section~\ref{sec:con}.

\section{Boussinesq equations} \label{sec:gov}
The dimensionless form of the two-dimensional (2D) incompressible Boussinesq equations can be written as follows:
\begin{align}
   \frac{\partial u}{\partial x} + \frac{\partial v}{\partial y} &=0,\label{eq:1} \\
   \frac{\partial u}{\partial t} + u\frac{\partial u}{\partial x} + v\frac{\partial u}{\partial y} &= - \frac{\partial P}{\partial x} + \frac{1}{Re}\Big(\frac{\partial^2 u}{\partial x^2} + \frac{\partial^2 u}{\partial y^2}\Big),\label{eq:2}\\
   \frac{\partial v}{\partial t} + u\frac{\partial v}{\partial x} + v\frac{\partial v}{\partial y} &= - \frac{\partial P}{\partial y} + \frac{1}{Re}\Big(\frac{\partial^2 v}{\partial x^2} + \frac{\partial^2 v}{\partial y^2}\Big) + Ri \theta , \label{eq:3}\\
   \frac{\partial \theta}{\partial t} + u\frac{\partial \theta}{\partial x} + v\frac{\partial \theta}{\partial y} &=   \frac{1}{Re Pr}\Big(\frac{\partial^2 \theta}{\partial x^2} + \frac{\partial^2 \theta}{\partial y^2}\Big) , \label{eq:4}
\end{align}
where $u$, $v$, $P$ and $\theta$ represent the horizontal and vertical components of the velocity field, pressure and temperature, respectively. Also, $Re$, $Pr$ and $Ri$ are Reynolds number (the ratio of viscous effects to inertial effects), Prandtl number (the ratio of the kinematic viscosity to the heat conductivity) and Richardson number (the ratio of the buoyancy force to the inertial forces), respectively.

Primitive variable formulation presented in Eqs.~\ref{eq:1}--\ref{eq:4} can be replaced by using the definition of the vorticity vector $\vec{\omega} = \nabla \times \boldsymbol u$ and the stream function $\psi$. Since we study the flow in a 2D domain, we only consider the $z$ component of $\vec{\omega}$, simply denoted as $\omega$. Therefore, the governing equations for the 2D incompressible Boussinesq equations can be rewritten in a dimensionless form as follows:

\begin{align}
   \frac{\partial \omega}{\partial t} + \frac{\partial \psi}{\partial y} \frac{\partial \omega}{\partial x} - \frac{\partial \psi}{\partial x} \frac{\partial \omega}{\partial y} &= \frac{1}{Re} \Big(\frac{\partial^2 \omega}{\partial x^2} + \frac{\partial^2 \omega}{\partial y^2}\Big) + Ri\frac{\partial \theta}{\partial x}, \label{eq:5} \\
   \frac{\partial \theta}{\partial t} + \frac{\partial \psi}{\partial y} \frac{\partial \theta}{\partial x} - \frac{\partial \psi}{\partial x} \frac{\partial \theta}{\partial y} &= \frac{1}{Re Pr} \Big(\frac{\partial^2 \theta}{\partial x^2} + \frac{\partial^2 \theta}{\partial y^2}\Big), \label{eq:6}
\end{align}
in which the vorticity and the stream function are coupled by the following kinematic relationship:
\begin{align}
   \frac{\partial^2 \psi}{\partial x^2} + \frac{\partial^2 \psi}{\partial y^2} = - \omega. \label{eq:7}
\end{align}
The flow velocity components can be recovered from the stream function using the following definitions:
\begin{align}
   u = \frac{\partial \psi}{\partial y},\;\;\;\;\; v = - \frac{\partial \psi}{\partial x}. \label{eq:8}
\end{align}
Here, we note that this vorticity and stream function formulation eliminates the pressure term from the Boussinesq equations, and hence, reduces the computational burden in two dimensional settings.

\section{Deterministic dynamic mode decomposition (DDMD)} \label{sec:ddmd}
DMD is a reduced order modeling method that aims to identify the spatiotemporal patterns of a dynamical system based on Koopman theory. In fact, this tool identifies the modal patterns underlying the phenomenon using spatiotemporal information that appears as a matrix whose rows and columns are the spatial coordinates and snapshots, respectively. The feature that makes DMD so special and easy to implement is that this reduced order modeling technique is completely data-driven that avoids any dependency on the physical knowledge of the system.

Vector $\mathbf{x} \in \mathbb{R}^n$ (where $n \gg 1$) is defined as the state of the dynamical system evolving in time as $\dfrac{d \mathbf{x}}{dt} = f(\mathbf{x},t)$, where $f(\cdot)$ can be any function depending on the dynamical nature of the problem. While the simplest form of the function is linear which is easy to be estimated in time, the challenge begins when non-linearity emerges in this function. The role of DMD here is to estimate the $\mathbf{A}$ operator, which converts the system equations from nonlinear to linear. This approximation is done by identifying the leading eigenvalues and the corresponding eigenvectors of the best fit linear operator $\mathbf{A}$ such that:
\begin{equation} \label{eq:cont_DMD_op}
    \frac{d \mathbf{x}}{dt} \approx \mathbf{A} \mathbf{x}.
\end{equation}

By estimating the spectral characteristics of the operator $\mathbf{A}$, which reflects the evolution of the system in different spatial coordinates, the DMD model can be created to represent the variation of the system through time. Therefore, the value of each spatial coordinate in the system is a combination of the DMD modes with different amplitudes along with the mode's decaying or growing rate at that special node of the system. There are also frequencies corresponding to each mode which defines its oscillation rate. In many applications, especially in fluid dynamics, the system spatial dimension $n$ should be chosen very large in order to enable the computational grid to be a proper representative of the actual flow field. Therefore, dimensionality reduction is a necessity for such models. One of the most common dimensionality reduction methods in DMD based ROM models is to obtain the most influential and powerful modes and truncate the rest of the modes. 

We start by defining the state matrix $\mathbf{X}$ as:
\begin{equation} \label{eq:full_data_mat}
 \mathbf{X} = \begin{bmatrix}
 | & | &   & | \\
 \mathbf{x}_{1} & \mathbf{x}_{2} & \dots & \mathbf{x}_{m} \\ 
 | & | &  & | \\
 \end{bmatrix},
\end{equation}
in which the information of $\mathbf{X} \in \mathbb{R}^{n \times m}$ is collected at different time steps $t_k = [t_1,t_2,\dots,t_m] \in \mathbb{R}^m$ where $n$ is the number of degrees of freedom and $m$ is the number of snapshots. Due to the time varying nature of the dynamical problems, the collected data is very similar to a movie that can be either obtained from experiments using sensors or from high fidelity numerical computations of differential equations governing the phenomenon. In this study, the data is obtained by solving the Boussinesq equations governing the Marsigli flow problem in which fluids with two different temperatures instantaneously meet each others by eliminating the barrier between them. This is denoted as the full order model (FOM) solution which is our reference of comparison for the rest of the paper. The problem is defined in 2D domain in the $x$ and $y$ directions, with $n_x$ and $n_y$ number of nodes, respectively. Thus, there is one $n_x \times n_y$ data matrix (e.g., temperature field for the present study) corresponding to each time step. Since the final data set which is supposed to contain the data related to every spatial coordinate at all time steps, has three dimensions (two spatial and one temporal dimensions), it is worth computationally to transform it into a 2D matrix by rearranging its spatial dimension in a column vector $\mathbf{x}^{(k)}\in \mathbb{R}^{n_x \cdot n_y}$ corresponding to a particular time $t_k$. Thus, the full data matrix $\mathbf{X} \in \mathbb{R}^{n \times m}$ is formed as Eq.~\ref{eq:full_data_mat}. Now, we can rewrite Eq.~\ref{eq:cont_DMD_op} in the discrete-time system sampled at $\Delta {t}$ in time as follows:
\begin{equation} \label{eq:dist_DMD_op}
    \mathbf{x}_{k+1} \approx \hat{\mathbf{A}} \mathbf{x}_{k}.
\end{equation}

Matrix $\hat{\mathbf{A}}$ should be estimated in a way wherein its eigenvectors and eigenvalues satisfy Eq.~\ref{eq:dist_DMD_op}. Thus, the full data set is split into two matrices $\mathbf{X}_1 \in \mathbb{R}^{n \times (m-1)},$ and $\mathbf{X}_2 \in \mathbb{R}^{n \times (m-1)}$ as defined below:
\begin{equation} \label{eq:split_data_mat}
 \mathbf{X}_1 = \begin{bmatrix}
 | & | &   & | \\
 \mathbf{x}_{1} & \mathbf{x}_{2} & \dots & \mathbf{x}_{m-1} \\ 
 | & | &  & | \\
 \end{bmatrix}, \qquad
 \mathbf{X}_2 = \begin{bmatrix}
 | & | &   & | \\
 \mathbf{x}_{2} & \mathbf{x}_{3} & \dots & \mathbf{x}_{m} \\ 
 | & | &  & | \\
 \end{bmatrix}.
\end{equation}
As a result, Eq.~\ref{eq:dist_DMD_op} can be rewritten as follows in matrix format:
\begin{equation} \label{eq:DMD_op}
    \mathbf{X}_2 \approx \hat{\mathbf{A}} \mathbf{X}_1,
\end{equation}
and the least squares optimization can be used to compute the operator $\hat{\mathbf{A}}$:
\begin{equation} \label{eq:optimization}
    \hat{\mathbf{A}} = \underset{\hat{\mathbf{A}}}{\arg\min} \|\mathbf{X}_2-\hat{\mathbf{A}}\mathbf{X}_1 \|_F \;,
\end{equation}
where $\|.\|_F$ is the Frobenius norm. Since $\hat{\mathbf{A}} \in \mathbb{R}^{n \times n}$, solving this optimization problem directly is computationally prohibitive because $n$ is very large in most of the desired applications in fluid dynamics. So, the objective is to replace this big matrix with lower rank approximations. 

In the standard form of the DMD, the linear operator $\hat{\mathbf{A}}$ is projected onto a lower $R$-dimensional subspace to reduce the computing cost of solving the optimization problem in Eq.~\ref{eq:optimization}. The singular value decomposition (SVD) of the matrix $\mathbf{X}_1$ as $\mathbf{X}_1 = \mathbf{U}\boldsymbol{\Sigma}\mathbf{V}^*$ (where $\mathbf{V}^*$ denotes the complex conjugate transpose of matrix $\mathbf{V}$) can be used to obtain suitable projection basis functions. The compact version of SVD can be derived in a way that \textcolor{rev1}{$\mathbf{X}_1 = \mathbf{U}^{}_\mathbf{R} \boldsymbol{\Sigma}^{}_\mathbf{R} \mathbf{V}^*_\mathbf{R}$}, where $\mathbf{U}^{}_\mathbf{R} \in \mathbb{R}^{n \times R}$ and $\mathbf{V}^{}_\mathbf{R} \in \mathbb{R}^{(m-1) \times R}$ represent the matrix of the first $R$ columns of $\mathbf{U}$ and $\mathbf{V}$ respectively, while $\boldsymbol{\Sigma}^{}_\mathbf{R} \in \mathbb{R}^{R \times R}$ is the first $R \times R$ dimensional sub-block of $\boldsymbol{\Sigma}$, with $R$ being the rank of $\boldsymbol{\Sigma}$. The projected $\hat{\mathbf{A}}$ onto the $R$-dimensional space is taken as:
\begin{equation} \label{eq:Atilde}
    \Tilde{\mathbf{A}} = \mathbf{U}^*_\mathbf{R} \hat{\mathbf{A}} \mathbf{U}^{}_\mathbf{R}.
\end{equation}

Hence, rearranging the optimization problem in Eq.~\ref{eq:optimization} gives:
\begin{equation}
    \Tilde{\mathbf{A}} = \underset{\Tilde{\mathbf{A}}}{\arg\min} \| \mathbf{X}_2-\mathbf{U^{}_R} \Tilde{\mathbf{A}} \boldsymbol{\Sigma}^{}_\mathbf{R} \mathbf{V}^*_\mathbf{R} \|_F,
\end{equation}
which leads to $\Tilde{\mathbf{A}} = \mathbf{U^*_R} \mathbf{X}_2 \mathbf{V^{}_R} \boldsymbol{\Sigma}^{-1}_{\mathbf{R}} \in \mathbb{R}^{R \times R}$. The eigenpairs of the constructed $\Tilde{\mathbf{A}}$ are in fact the eigenvalues and the eigenvectors of the DMD and can be calculated using eigenvalue decomposition of $\Tilde{\mathbf{A}}$ as $\Tilde{\mathbf{A}} \mathbf{W} = \mathbf{W} \boldsymbol{\Lambda}$. The eigenvalue decomposition of $\Tilde{\mathbf{A}}$ may result in complex values in either $\mathbf{W}$ or $\boldsymbol{\Lambda}$, which is an implication of the existence of oscillatory modes in the system. The columns of matrix $\mathbf{W} \in \mathbb{C}^{R \times R}$ are eigenvectors while the diagonal elements of $\boldsymbol{\Lambda} = \text{diag} (\{ \lambda_i \}_{i=1}^R) \in \mathbb{C}^{R \times R}$ represent the eigenvalues. Finally, the DMD modes can be computed in the following way: 
\begin{equation}
    \boldsymbol{\Phi} = \mathbf{U}_\mathbf{R}\mathbf{W}.
\end{equation}

\begin{algorithm}[ht!]
\caption{Deterministic Dynamic Mode Decomposition (DDMD)}
\label{alg:std_DMD}
\begin{algorithmic}[1]
\STATE The matrix $\mathbf{X}$ is split into two matrices $\mathbf{X}_1 = \{ \mathbf{x}_{1}, \mathbf{x}_{2}, \dots, \mathbf{x}_{m-1} \}$ and $\mathbf{X}_2 = \{ \mathbf{x}_{2}, \mathbf{x}_{3}, \dots, \mathbf{x}_{m} \}$. \\
\vspace{10pt}
\STATE Perform SVD on $\mathbf{X}_1$  \vspace{-10pt}
\begin{align*}
    \mathbf{U} \boldsymbol{\Sigma} \mathbf{V}^* = \text{svd} {\left( \mathbf{X}_1 \right)}
\end{align*}

\STATE Rank truncation [to reduce noise] \vspace{-10pt}
\begin{align*}
    \mathbf{U}_\mathbf{R} &= \mathbf{U}(:,1:R) \\
    \mathbf{V}_\mathbf{R} &= \mathbf{V}(:,1:R) \\
    \boldsymbol{\Sigma}_\mathbf{R} &= \boldsymbol{\Sigma}(1:R,1:R)
\end{align*}

\STATE Low-rank dynamics \vspace{-10pt}
\begin{align*}
    \Tilde{\mathbf{A}} = \mathbf{U^*_R} \mathbf{X}_2 \mathbf{V}^{}_\mathbf{R} \boldsymbol{\Sigma}^{-1}_\mathbf{R}
\end{align*}

\STATE Eigenvalue decomposition \vspace{-10pt}
\begin{align*}
    \left[\mathbf{W}, \boldsymbol{\Lambda} \right] = \text{eig} {( \Tilde{\mathbf{A}} )}
\end{align*}

\STATE Compute DMD modes and spectrum \vspace{-10pt}
\begin{align*}
\boldsymbol{\Psi} &= \mathbf{U}_\mathbf{R} \mathbf{W} \quad \text{or} \quad \boldsymbol{\Psi} = \mathbf{X}_2 \mathbf{V}^{}_\mathbf{R} \boldsymbol{\Sigma}^{-1}_\mathbf{R} \mathbf{W} \\
\lambda_i &= \{ \text{diag} (\boldsymbol{\Lambda}) \} \\
\alpha_i &= \text{ln} {(\lambda_i)}/\Delta T
\end{align*}
\end{algorithmic}
\end{algorithm}

There is also another way of constructing $\boldsymbol{\Phi}$. This type is called ``exact'' DMD, which is the procedure implemented in the present study \citep{kutz2016dynamic}:
\begin{equation}
    \boldsymbol{\Phi} = \mathbf{X}_2 \mathbf{V}^{}_\mathbf{R} \boldsymbol{\Sigma}^{-1}_\mathbf{R} \mathbf{W}.
\end{equation}
Each column of $\boldsymbol{\Phi}$ represents a mode of the system that is used to construct the field at every node of the system. On the other hand, the contribution of these modes in the construction of the field is determined by a weight matrix which includes eigenvectors $\mathbf{W}$ and the eigenvalues $\Lambda$. In this regard, one can simply truncate the modes with relatively lower weights. This restricts the solution to a reduced order approximation $\boldsymbol{\Phi}^{}_\mathbf{r}$ which contains the most influential $r$ columns of $\boldsymbol{\Phi}$ to form the reduced DMD mode matrix $\boldsymbol{\Phi}_\mathbf{r} = \{\boldsymbol{\phi}_i\}_{i=1}^r \in \mathbb{C}^{n \times r}$. Similarly, the eigenvalues corresponding to the retained $r$ modes create a new reduced diagonal eigenvalue matrix $\boldsymbol{\Lambda}_\mathbf{r} = \text{diag}\{ \lambda_i \}_{i=1}^{r} \in \mathbb{C}^{r \times r}$. At the end of this step, the dominant spatial patterns of the field are recognized and associated with the eigenvalues that now should be defined in continuous-time expression as:
\begin{equation}\label{eq:dis_cont_evalue}
    \textcolor{rev1}{\alpha_i} = \frac{\ln{\lambda_i} }{\Delta t}.
\end{equation}

\textcolor{rev1}{Algorithm ~\ref{alg:std_DMD} presents a summary of the algorithmic processes for the deterministic DMD.}
As introduced in the discretizing step (Eq.~\ref{eq:dist_DMD_op}), $\Delta t$ is the time interval between two consecutive snapshots and the vector $\boldsymbol{\textcolor{rev1}{\alpha}} = \{ \textcolor{rev1}{\alpha_i} \}_{i=1}^{r} \in \mathbb{C}^{r}$ contains continuous-time eigenvalues. Providing all the requirements, it is possible to reconstruct the high dimensional dynamics from the obtained reduced order components as follows: 
\begin{equation} \label{eq:recon_dis}
    \mathbf{x}_{k}^{ROM} = \sum_{i=1}^{r}\boldsymbol{\phi}_i \lambda_i^{k-1} b_i = \boldsymbol{\Phi}^{}_\mathbf{r} \boldsymbol{\Lambda}^{k-1}_\mathbf{r} \mathbf{b},
\end{equation}
where $\mathbf{x}_{k}^{ROM}$ is the field in $k$th snapshot and $b_i$ is the initial amplitude of each mode and $\mathbf{b} = \{ b_i \}_{i=1}^{r} \in \mathbb{C}^{r}$ is the vector of initial amplitudes of the DMD modes given as:
\begin{equation} \label{eq:amp}
     \mathbf{b} = \boldsymbol{\Phi}^{\dagger}_\mathbf{r} \mathbf{x}_{1},
\end{equation}
where $\boldsymbol{\Phi_{r}^{\dagger}}$ is the Moore-Penrose pseudoinverse of $\boldsymbol{\Phi}^{}_\mathbf{r}$ and $x_1$ is the first column (the first snapshot or the initial condition of the system) of $\mathbf{X}_1$ (defined in Eq.~\ref{eq:split_data_mat}). Finally, by converting the discrete eigenvalues to their continuous version (Eq.~\ref{eq:dis_cont_evalue}), the dynamics can be expressive for any time in the domain as:
\begin{equation} \label{eq:recon_cont}
     \mathbf{x}_{k}^{ROM} = \sum_{i=1}^{r} \boldsymbol{\phi}_i e^{\textcolor{rev1}{\alpha_i} t_k} b_i = \boldsymbol{\Phi}^{}_\mathbf{r} \text{diag}[e^{\boldsymbol{\textcolor{rev1}{\alpha}}t_k}] \mathbf{b}.
\end{equation}

In Eq.~\ref{eq:recon_cont}, the initial amplitude vector $\mathbf{b}$ determines the value by which each spatial coordinate (node) starts its variation. The variation dictates to each node spatially by the DMD modes $\phi_i$ and temporally by $\textcolor{rev1}{\alpha_i}$. The field evolves through time using continuous-time eigenvalues, the real component of which determines the mode's growth/decay rate. A positive real part's mode develops over time, while a negative real part's mode decays. The imaginary portion, on the other hand, determines the mode's oscillation frequency.

\section{Randomized dynamic mode decomposition (RDMD)} \label{sec:rdmd}
Underneath every high-dimensional data, there are lower-dimensional patterns that regulate the majority of the dynamics. This is the main concept of ROM that motivates compression of the data. Randomized dynamic mode decomposition (RDMD) is a version of DMD that utilizes this idea in a nonintrusive sense. Instead of working with the whole data set, the RDMD method exploits special parts of the data matrix called the sketch which is just another matrix that is substantially smaller than the original system but nevertheless properly approximates it. Such sketching or embedding is done by applying random sampling of the input matrix with specific features to produce a compressed version of the original system. The time-consuming calculations are thus performed on the sketch, followed by post-processing to map the outputs back to the original space. There are a few sketching-based algorithms which have been implemented in some fluid dynamics problems such as sketching the range of $\mathbf{X}_1$, sketching the range of $\mathbf{X}$, and sketching the range and the corange of $\mathbf{X}_1$ \citep{bistrian2017randomized, ahmed2022sketching, ahmed2022dynamic}. In the present study, we only consider sketching range of $X_1$ framework.

In this randomized DMD architecture which, proposed by \citet{bistrian2017randomized}, the goal of lowering the SVD cost for the data matrix $\mathbf{X}_1 \in \mathbb{R}^{n \times m-1}$ is achieved by generating a near-optimal basis with a target rank of $k$ using random projections. This projection is aimed to capture the range of $\mathbf{X}_1$ while producing a smaller sketch matrix $\mathbf{X}_1^{red} \in \mathbb{R}^{k \times m-1}$. 
This randomization is expected to improve the efficiency of later deterministic model reduction processes in terms of computational time and memory requirements. More specifically, to find the lowered rank matrix $\mathbf{X}_1^{red}$, we first perform a randomized projection of the input data matrix $\mathbf{X_1}$ as follows:
\begin{equation} \label{eq:rand_X1}
     \mathbf{X}_1^{rand} = \mathbf{X}_1 \Omega,
\end{equation}
where $\mathbf{X}_1^{rand}\in \mathbb{R}^{n \times k}$ is the randomly sampled matrix and $\mathbf{\Omega} \in \mathbb{R}^{(m-1) \times k}$ is a randomized matrix derived from a Gaussian distribution that integrates the concept of randomness. Now, $k$, which denotes the desired rank, should be defined as an adjustable parameter. In RDMD, this target rank is defined as $k=r+s$, where $r$ represents the number of eventually retained modes and $s$ is an oversampling factor that aids in obtaining a better basis. In this study, we use $s=10$ for all reduced order models. \textcolor{rev1}{Typically, a larger value of $s$ yields a better approximation of the data matrix. However, it increases the computational cost of the algorithm. Although the results might vary when the factors such as $s$ and $r$ (which affect the dimension of matrix $\mathbf{X}_1^{rand}$) change, we found that an oversampling factor $s=10$ gives a reasonable accuracy-cost trade-off for $O(10)$ DMD modes.}

Then, an orthonormal basis of the derived $\mathbf{X_1}^{rand}$ is created using QR decomposition of this random selected matrix:
\begin{equation} \label{eq:QR_X1(rand)}
     \mathbf{X}_1^{rand} = \mathbf{Q} \mathbf{R},
\end{equation}
where $\mathbf{Q} \in \mathbb{R}^{n \times k}$ is an orthonormal matrix and $\mathbf{R} \in \mathbb{R}^{k \times k}$ is an upper triangular matrix that is not required in this procedure. Next, to obtain a lower dimension matrix $\mathbf{X_1}^{red}\in \mathbb{R}^{k \times (m-1)}$, the full order data $\mathbf{X_1}$ is projected onto the basis $\mathbf{Q}$ as follows:
\begin{equation} \label{eq:X1_reduced}
     \mathbf{X}_1^{red} = \mathbf{Q^*} \mathbf{X}_1,
\end{equation}
where $\mathbf{Q^*}$ denotes the conjugate transpose of $\mathbf{Q}$. Now, the randomization strategy is implemented into the data and SVD can be simply performed on the reduced version of data matrix as $\mathbf{X}_1^{red} = \Tilde{\mathbf{U}} \Tilde{\mathbf{\Sigma}} \Tilde{\mathbf{V}^*}$. Finally, the SVD components of the full data matrix $\mathbf{X}_1$ can be recovered from their reduced peers as follows:
\begin{align} 
     \mathbf{U} &= \mathbf{Q} \Tilde{\mathbf{U}}, \label{eq:U_recover} \\
     \mathbf{\Sigma} &= \Tilde{\mathbf{\Sigma}},  \label{eq:S_recover} \\
     \mathbf{V} &= \Tilde{\mathbf{V}}.  \label{eq:V_recover}
\end{align}

\begin{algorithm}[ht!]
\caption{Randomized Dynamic Mode Decomposition (RDMD) by Sketching the Range of $\mathbf{X}_1$}
\label{alg:bist}
\begin{algorithmic}[1]
\STATE The matrix $\mathbf{X}$ is split into two matrices $\mathbf{X}_1 = \{ \mathbf{x}_{1}, \mathbf{x}_{2}, \dots, \mathbf{x}_{m-1} \}$ and $\mathbf{X}_2 = \{ \mathbf{x}_{2}, \mathbf{x}_{3}, \dots, \mathbf{x}_{m} \}$. \\
\vspace{10pt}
\STATE Draw a random matrix $\boldsymbol{\Omega}_1 \in \mathbb{R}^{(m-1) \times k}$ from Gaussian distribution and perform the randomized projection of $\mathbf{X}_1$ \vspace{-10pt}
\begin{align*}
    \mathbf{X}_1^{rand} = \mathbf{X}_1 \boldsymbol{\Omega}
\end{align*}

\STATE Perform QR decomposition as $\mathbf{X}_1^{rand}=\mathbf{Q}\mathbf{R}$ to obtain a near-optimal basis $\mathbf{Q}$ for $\mathbf{X}_1$ and discard $\mathbf{R}$.
\vspace{10pt}
\STATE A sketch $\mathbf{B}$ of $\mathbf{X}_1$ is obtained as  \vspace{-10pt}
\begin{align*}
        \mathbf{X}_1^{red} = \mathbf{Q}^{*} \mathbf{X}_1
\end{align*}

\STATE Perform SVD on $\mathbf{X}_1^{red}$ \vspace{-10pt}
\begin{align*}
    \tilde{\mathbf{U}} \tilde{\boldsymbol{\Sigma}} \tilde{\mathbf{V}}^* = \text{svd} {\left( \mathbf{X}_1^{red} \right)}
\end{align*}

\STATE Recover SVD of $\mathbf{X}_1$ \vspace{-10pt}
\begin{align*}
    \mathbf{U} &= \mathbf{Q} \tilde{\mathbf{U}} \\
    \boldsymbol{\Sigma} &= \tilde{\boldsymbol{\Sigma}} \\
    \mathbf{V} &=  \tilde{\mathbf{V}}
\end{align*}

\STATE Rank truncation [to reduce noise] \vspace{-10pt}
\begin{align*}
    \mathbf{U}_\mathbf{R} &= \mathbf{U}(:,1:R) \\
    \mathbf{V}_\mathbf{R} &= \mathbf{V}(:,1:R) \\
    \boldsymbol{\Sigma}_\mathbf{R} &= \boldsymbol{\Sigma}(1:R,1:R)
\end{align*}

\STATE Low-rank dynamics \vspace{-10pt}
\begin{align*}
    \Tilde{\mathbf{A}} = \mathbf{U^*_R} \mathbf{X}_2 \mathbf{V}^{}_\mathbf{R} \boldsymbol{\Sigma}^{-1}_\mathbf{R}
\end{align*}

\STATE Eigenvalue decomposition \vspace{-10pt}
\begin{align*}
    \left[\mathbf{W}, \boldsymbol{\Lambda} \right] = \text{eig} {( \Tilde{\mathbf{A}} )}
\end{align*}

\STATE Compute DMD modes and spectrum \vspace{-10pt}
\begin{align*}
\boldsymbol{\Psi} &= \mathbf{U}_\mathbf{R} \mathbf{W} \quad \text{or} \quad \boldsymbol{\Psi} = \mathbf{X}_2 \mathbf{V}^{}_\mathbf{R} \boldsymbol{\Sigma}^{-1}_\mathbf{R} \mathbf{W} \\
\lambda_i &= \{ \text{diag} (\boldsymbol{\Lambda}) \} \\
\alpha_i &= \text{ln} {(\lambda_i)}/\Delta T
\end{align*}
\end{algorithmic}
\end{algorithm}  

\textcolor{rev1}{Algorithm~\ref{alg:bist} summarizes the algorithmic steps for the randomized DMD based on sketching the range of $\mathbf{X}_1$.}

\section{Nonlinear proper orthogonal decomposition (NLPOD)} \label{sec:nlpod}
\citet{ahmed2021nonlinear} have proposed the nonlinear proper orthogonal decomposition (NLPOD) framework which supersedes the Galerkin projection (GP). Unlike the GP method that requires the access to the partial differential equations describing the desired physical system, the NLPOD model can be built using an autoencoder-based method that does not rely on prior knowledge of the underlying equations. The procedure starts with the assumption that the desired field, which is temperature field in our demonstration example, can be estimated properly using a large, but finite, number of orthonormal functions existing in a predetermined collection of basis functions derived from POD modes. Like what we utilize in DMD, a data matrix $\mathbf{X} \in \mathbb{R}^{n \times m}$ is defined in which the columns are the snapshots collected from spatial coordinates. 

For performing POD, we can simply apply SVD to the data set as $\mathbf{X}=\mathbf{U} \mathbf{\Sigma} \mathbf{V^*}$ and use the columns of the left singular matrix $\mathbf{U}$ as the POD basis set, $\mathbf{U}=[\mathbf{\phi}_1, \mathbf{\phi}_2, ...]$. The matrix $\mathbf{\Sigma}$ can also be used to determine the number of the required basis functions (modes) to build the POD subspace as this matrix is a representation of the energy content in the corresponding modes. Thus, the orthonormal basis set should have an adequate number of modes to be able to cover an acceptable relative information content (RIC) defined as follows:
\begin{equation} \label{eq:RIC}
     \mathbf{RIC(\%)} = \frac{\sum_{i=1}^{r} {{\sigma_i}^2}}{\sum_{i=1}^{m} {{\sigma_i}^2}}\times100 , 
\end{equation}
where $\mathbf{\sigma_i}$ is the $i$th singular value and $r$ is the number of selected modes to reconstruct the field. We note that $r$ is defined to be less than or equal to $m$, implying that the maximum number of available modes is equal to the number of collected snapshots. 

Now, the field can be estimated using the combination of the chosen $r$ modes as:
\begin{equation} \label{eq:r-ranked field}
     \mathbf{X}(\mathbf{x}, t) = \sum_{i=1}^{r} {a_i(t) \mathbf{\phi}_i (x)},
\end{equation}
where $a_i(t)$ is the coefficient of the $i$th column of $U$. These coefficients can be found by substituting $\mathbf{X}(\mathbf{x}, t)$ introduced in Eq.~\ref{eq:r-ranked field} into Eq.~\ref{eq:6} as the temperature $\theta$. Since $\mathbf{\phi}_i (x)$ are orthonormal vectors, the coefficients $a_i(t)$ are derived by simply performing the dot product of the corresponding $\mathbf{\phi}_i (x)$ into the equations. Exploiting the orthogonality properties, all the terms in the Eq.~\ref{eq:r-ranked field} will get zero except the one we are looking for its coefficient as follows: 
\begin{equation} \label{eq:coeff_solution}
     \mathbf{a} = \mathbf{\Phi}^T \mathbf{X},
\end{equation}
where $\mathbf{\Phi}^T(x)$ is the transpose of $\mathbf{\Phi} \in R^{n \times r}$ which is the same as matrix $U_r$ obtained by choosing $r$ columns of $U$ derived from the SVD. Next, the driven coefficients are fed into a nonlinear neural network AE to be mapped to a low dimensional latent space $z$ and then decode it back to the original dimension at the output, with the goal of minimizing the reconstruction loss $\mathcal{L}=|\mathbf{a}-\Tilde{\mathbf{a}}|^2$ where $\Tilde{\mathbf{a}}$ is the reconstructed data using the AE \citep{pan2021sparsity}. The manifold learning can be described by autoencoder as follows (using the encoder function as $\zeta$ and the decoder function as $\eta$):
\begin{equation} \label{eq:AE_encoder}
    \mathbf{a}(t) \in R^r \xrightarrow[]{\mathbf{\eta}} \mathbf{z}\in R^l, 
\end{equation}
\begin{equation} \label{eq:AE_decoder}
    \mathbf{z}\in R^l \xrightarrow[]{\mathbf{\zeta}}  \mathbf{a}(t) \in R^r, 
\end{equation}
where $l$ denotes the dimension of the latent variable existing in $\mathbf{z}$.

A surrogate model emulator is built to evolve the latent space variables onto the manifold revealed by the AE for temporal dynamics. The capabilities of LSTM networks in sequential data prediction are used in this study to transmit the latent variables in time. Thus a trained network that can estimate the coefficient in latent space $\mathbf{z}$ is available. Then, the estimated compressed variables $\mathbf{z}$ is passed through the decoder to provide the coefficients $\mathbf{a}$ of the POD expansion. Finally, the sought field can be reconstructed as the modes and their coefficients are available. \textcolor{rev1}{Table~\ref{table:hyper} represents the value of hyperparameters used for running the networks.} 
\vspace{20pt}

\setlength{\tabcolsep}{5pt} 
\renewcommand{\arraystretch}{1.5} 
\begin{table*}[htbp!]
\caption{\textcolor{rev1}{Hyperparameters used in autoencoder (AE) and long short-term memory (LSTM) networks}. } \vspace{5pt}
\centering
\fontsize{6.5}{12}
\begin{tabular}{ |p{4.1cm}|p{4.6cm}|p{4.6cm}|  }
\hline
Parameters & AE & LSTM \\
\hline
Number of hidden layers & 4 in encoder (128, 64, 32, 8) 4 in decoder (8, 32, 64, 128) & 3 (each with 10 units) \\
Batch size & 32 & 32 \\
Epochs & 200 & 200 \\
Activation function & tanh  & tanh  \\
Activation function at the last layer & linear & linear \\
Validation-training ratio & 20\% & 20\% \\
Loss function & MSE & MSE   \\
Optimizer & Adam & Adam \\
\hline
\end{tabular}
\label{table:hyper}
\end{table*}

\section{Numerical results} \label{sec:res}
In this section, the performance of the described ROM methods including DDMD, RDMD and NLPOD is compared through the reconstruction of the temperature field existing in the Boussinesq equations. More specifically, the temperature field is obtained from the Marsigli flow problem, where a fluid is divided into two partitions with differing temperatures. The dividing barrier is immediately removed, allowing the fluids to slide over each other in a convection- and buoyancy-driven manner. Readers are referred to \citet{san2015principal, ahmed2021multifidelity} for further details about the computational setup for this problem. 

The full order model (FOM) is obtained by solving the partial differential equations (PDE) in a two dimensional domain with a uniform Cartesian grid of dimension $512 \times 64$.
The vorticiy is set as zero on the boundaries imposing free-slip boundary conditions. Also, an adiabatic boundary condition for temperature is applied on the walls. \textcolor{rev1}{To initiate the thermal interaction, the defined two dimensional geometry is divided into two parts. We set the initial temperature to 1.5 for all nodes on the left of the domain while the initial temperature of the right side of the frame is set to 1.} 
In FOM, the temperature of the fluid is measured through an 8-second time period with a time step of $5 \times 10^{-4}$ second intervals to achieve the high fidelity data. Also, the dimensionless parameters appearing in the Boussinesq Eqs.~\ref{eq:5}--\ref{eq:7} are defined as follows: Re $=1000$, Ri $=4$, and Pr $=1$.

Among the available spatiotemporal data provided by FOM solution, 200 equally spaced snapshots in the same 8-second time period are used to train the estimator models either in the case of DMD methods or NLPOD. The reconstructed temperature field using 10 modes is shown at $t=5$ and $t=8$ in Figure~\ref{fig:Fig1} for different ROM methods and FOM as well. \textcolor{rev1}{Our results indicate that the error is monotonically decreasing for increasing numbers of modes for $t=5$, whereas the contrary is the case for $t=8$. In the Marsigli flow dynamics, shear layer evolves and smaller scales develop gradually when we integrate in time. Therefore, the reconstruction becomes a more challenging task in later times. We highlight that time $t=5$ represents an instance before the shear front reaches the side walls. However, the shear fronts pass through the left and right side walls at time $t=8$. Therefore, boundary effects are more pronounced at $t=8$.} Moreover, it can be seen that the NLPOD performs more accurately than both DMD methods, especially on boundaries. Besides, RDMD has given smoother field particularly in the final snapshot. Table~\ref{table:comparison} presents the root mean squared error (RMSE) of the introduced reconstruction methods for different number of applied modes. As it is shown in Figure~\ref{fig:Fig2}, by increasing the number of modes to 20, the obtained fields from DDMD and RDMD become twice as accurate in terms of RMSE, while this improvement in error is around five fold in the case of NLPOD. Moreover, we observe that the boundary regions are more precisely reconstructed in all the cases compared to $r=10$. 

Figure~\ref{fig:Fig3} shows the RMSE of the introduced reconstructed models (with different number of modes) at two different times, $t=5$ (top) and $t=8$ (bottom). At the earlier time of the flow evolution (i.e., $t=5$) where there is a relatively fewer fine scale vortical structures, results indicate that both DMD approaches monotonically decrease errors as we increase the number of retained modes. \textcolor{rev1}{On the other hand, the behavior of the error of both DMD methods change for $t=8$, where the RMSE increases as the number of modes increases. The DMD models approximates the Koopman operator using different number of modes (Eq.~\ref{eq:Atilde}). Thus, by using more modes, the effect of weaker modes (modes corresponding to small eigenvalues in matrix $\Sigma$) is taken into account. In case of DMD, the order of nonlinearity of these low-amplitude modes may define the error propagation in both spatial and temporal domains (the rows and columns of matrix $X$).}

Figure~\ref{fig:Fig4} illustrates the temperature field using 50 modes. In spite of that the NLPOD surpasses the DMD methods in describing the boundaries, they perform equally well, especially in case of the final snapshot $t=8$. By continuing the trend of increasing the modes (Figure~\ref{fig:Fig5}), we get more accurate results with smoother edges that are able to reconstruct the boundaries in a physically meaningful way. It is worth to note that at the final time ($t=8$) where more fine vortical patterns are expected, DMD reconstruction error unexpectedly increases with increasing the number of modes. In contrast, we do not observe such monotonic behavior in the NLPOD approach. These observations can be related to the suboptimal selection of the relevant hyperparamaters that intensify overfitting issues. Nonetheless, the results indicate that all three approaches are in a good agreement with the full order model solution when a reasonable combination of hyperparameters (e.g., number of modes, oversampling factor, nueral network architecture, etc.) is selected.


\textcolor{rev1}{We emphasize that the purpose of our analysis is mainly two-fold: (i) to explore the size of data snapshots in building reduced order models which is performed by creating models using different number of POD modes, and (ii) to investigate the sensitivity of noise embedded in data snapshots. To address the second purpose, we add synthetic Gaussian white noise to the data. The outputs of the models are compared through different amplitudes of noise including: 2\%, 5\% and 10\%. Specifically, Figure~\ref{fig:Fig6}--\ref{fig:Fig9} depict the reconstructed field with the presence of 2\% noise while different number of modes are considered. The corresponding RMSE is presented in Figure~\ref{fig:Fig10} and Table~\ref{table:comparison2}. It can be observed that implementing more modes in the models improves the reconstruction in both early and final snapshots. Table~\ref{table:comparison5} and Table~\ref{table:comparison10} present the RMSE for adding 5\% and 10\% noise, respectively. In both DMD methods when the earlier snapshot is taken into consideration, if we compare the results in Tables~\ref{table:comparison}--\ref{table:comparison10} that use the same approach and the same number of modes, we can see that the RMSE grows as the noise amplitude increases. 
}

\textcolor{rev1}{Figure~\ref{fig:Fig11}--\ref{fig:Fig14} show the reconstructed field by using different number of modes while the noise amplitude is set to 5\%. As it can be seen, the higher noise level affects the reconstructed fields, especially on the boundaries where there are more variations in the temperature field. In all methods, the reconstruction improves by implementing more modes to the input. It is also observable that the NLPOD reconstructs the fields more efficiently than DMD methods. There is just an exception in this trend where the error increases as the number of modes is chosen $r=80$ (Figure~\ref{fig:Fig15}). This deviation from the main trend can be observed in the bottom row of Figure~\ref{fig:Fig14}, specially in $t=5$. Obviously, the field in $t=5$ is scaled in its developing direction and this can be the source of this relatively higher error in the case that the NLPOD is used. }

\textcolor{rev1}{To show the effect of noise on the reconstruction process, we repeat the process with adding 10\% noise to the input. The results are illustrated in Figure~\ref{fig:Fig16}--\ref{fig:Fig19} for different number of modes. Although there is not considerable differences between the construction errors resulting from different approaches, it seems that the NLPOD model is more powerful in damping the effect of the noise compared to the DMD methods (Figure~\ref{fig:Fig6}, Figure~\ref{fig:Fig11}, and Figure~\ref{fig:Fig16}). Comparing corresponding results in  Figures~\ref{fig:Fig11}--\ref{fig:Fig14} and Figures~\ref{fig:Fig16}--\ref{fig:Fig19} gives a brighter view of the effect of the noise on different methods. As the noise level increases, DMD methods require more modes to be able to attenuate the effect of the noise in the reconstructed field. Despite the noise's effect being observable in the results of NLPOD method, the performance of NLPOD model is better in simultaneously reconstructing the field and damping the noise of the input. }

\textcolor{rev1}{Figure~\ref{fig:Fig20} shows the error resulted from 10\% noise for different number of modes. It can be implied that DMD models behave more monotonically than NLPOD in terms of the reconstruction error. Moreover, as the number of modes changes, adding noise to both DMD approaches produces errors with smaller variations.}

\begingroup
\setlength{\tabcolsep}{5pt} 
\renewcommand{\arraystretch}{1.5} 
\begin{table*}[htbp!]
\caption{Reconstruction RMSE obtained exploiting various approaches (DDMD, RDMD and NLPOD) using different models with different number of modes ($r$) \textcolor{rev1}{by using 200 snapshots without synthetic noise}. } \vspace{5pt}
\centering
\fontsize{6.5}{12}
\begin{tabular}{ |p{0.95cm}||p{1.3cm}|p{1.3cm}|p{1.3cm}|p{1.3cm}||p{1.3cm}|p{1.3cm}|p{1.3cm}|p{1.3cm}| }
 \hline
  & \multicolumn{3}{c}{\small$t=5$} & & \multicolumn{3}{c}{\small$t=8$} & \\ \hline
 {\scriptsize Method} & \small$r=10$ & \small$r=20$ & \small$r=50$ & \small$r=80$ & \small$r=10$ & \small$r=20$ & \small$r=50$ & \small$r=80$ \\
 \hline 
 {\scriptsize DDMD} & $1.28 \times 10^{-3}$ & $5.91\times 10^{-4}$ & $1.97\times 10^{-4}$ & $9.7\times 10^{-5}$ & $4.51\times 10^{-2}$ & $4.73\times 10^{-2}$ & $4.87\times 10^{-2}$ & $4.92\times 10^{-2}$ \\
 {\scriptsize RDMD} & $1.48\times 10^{-3}$ & $8.48\times 10^{-4}$ & $2.33\times 10^{-4}$ & $9.3\times 10^{-5}$ & $4.5\times 10^{-2}$ & $4.67\times 10^{-2}$ & $4.86\times 10^{-2}$ & $4.92\times 10^{-2}$ \\
 {\scriptsize NLPOD} & $1.0\times 10^{-3}$ & $2.13\times 10^{-4}$ & $7.6\times 10^{-4}$ & $2.73\times 10^{-3}$ & $5.05\times 10^{-2}$ & $4.89\times 10^{-2}$ & $4.68\times 10^{-2}$ & $5.35\times 10^{-2}$ \\
 \hline
\end{tabular}
\label{table:comparison}
\end{table*}

\begingroup
\setlength{\tabcolsep}{5pt} 
\renewcommand{\arraystretch}{1.5} 
\begin{table*}[htbp!]
\caption{Reconstruction RMSE obtained exploiting various approaches (DDMD, RDMD and NLPOD) using different models with different number of modes ($r$) \textcolor{rev1}{by using 200 snapshots with 2\% synthetic noise}. } \vspace{5pt} \vspace{5pt}
\centering
\fontsize{6.5}{12}
\begin{tabular}{|p{0.95cm}||p{1.3cm}|p{1.3cm}|p{1.3cm}|p{1.3cm}||p{1.3cm}|p{1.3cm}|p{1.3cm}|p{1.3cm}| }
 \hline
  & \multicolumn{3}{c}{\small$t=5$} & & \multicolumn{3}{c}{\small$t=8$} & \\ \hline
 {\scriptsize Method} & \small$r=10$ & \small$r=20$ & \small$r=50$ & \small$r=80$ & \small$r=10$ & \small$r=20$ & \small$r=50$ & \small$r=80$ \\
 \hline 
 {\scriptsize DDMD} & $1.35 \times 10^{-3}$ & $6.62\times 10^{-4}$ & $2.78\times 10^{-4}$ & $1.87\times 10^{-4}$ & $4.52\times 10^{-2}$ & $4.73\times 10^{-2}$ & $4.87\times 10^{-2}$ & $4.91
\times 10^{-2}$ \\
 {\scriptsize RDMD} & $1.5 \times 10^{-3}$ & $9.43
\times 10^{-4}$ & $3.26 \times 10^{-4}$ & $2.09 \times 10^{-5}$ & $4.5 \times 10^{-2}$ & $4.67 \times 10^{-2}$ & $4.87\times 10^{-2}$ & $4.89\times 10^{-2}$ \\
 {\scriptsize NLPOD} & $1.12\times 10^{-3}$ & $2.49
\times 10^{-4}$ & $2.95 \times 10^{-4}$ & $9.1 \times 10^{-5}$ & $5.1 \times 10^{-2}$ & $4.85 \times 10^{-2}$ & $5.35 \times 10^{-2}$ & $4.9 \times 10^{-2}$ \\
 \hline
\end{tabular}
\label{table:comparison2}
\end{table*}

\begingroup
\setlength{\tabcolsep}{5pt} 
\renewcommand{\arraystretch}{1.5} 
\begin{table*}[htbp!]
\caption{Reconstruction RMSE obtained exploiting various approaches (DDMD, RDMD and NLPOD) using different models with different number of modes ($r$) \textcolor{rev1}{by using 200 snapshots with 5\% synthetic noise}. } \vspace{5pt} \vspace{5pt}
\centering
\fontsize{6.5}{12}
\begin{tabular}{|p{0.95cm}||p{1.3cm}|p{1.3cm}|p{1.3cm}|p{1.3cm}||p{1.3cm}|p{1.3cm}|p{1.3cm}|p{1.3cm}| }
 \hline
  & \multicolumn{3}{c}{\small$t=5$} & & \multicolumn{3}{c}{\small$t=8$} & \\ \hline
 {\scriptsize Method} & \small$r=10$ & \small$r=20$ & \small$r=50$ & \small$r=80$ & \small$r=10$ & \small$r=20$ & \small$r=50$ & \small$r=80$ \\
 \hline 
 {\scriptsize DDMD} & $1.73 \times 10^{-3}$ & $1.06 \times 10^{-3}$ & $7.01 \times 10^{-4}$ & $6.59 \times 10^{-4}$ & $4.55 \times 10^{-2}$ & $4.76\times 10^{-2}$ & $4.87\times 10^{-2}$ & $4.89\times 10^{-2}$ \\
 {\scriptsize RDMD} & $1.93\times 10^{-3}$ & $1.39\times 10^{-3}$ & $8.56 \times 10^{-4}$ & $7.57\times 10^{-4}$ & $4.53\times 10^{-2}$ & $4.69\times 10^{-2}$ & $4.86\times 10^{-2}$ & $4.86\times 10^{-2}$ \\
 {\scriptsize NLPOD} & $7.67\times 10^{-4}$ & $3.14\times 10^{-4}$ & $1.02 \times 10^{-4}$ & $1.4 \times 10^{-3}$ & $4.92\times 10^{-2}$ & $4.99\times 10^{-2}$ & $4.9\times 10^{-2}$ & $6.22\times 10^{-2}$ \\
 \hline
\end{tabular}
\label{table:comparison5}
\end{table*}

\begingroup
\setlength{\tabcolsep}{5pt} 
\renewcommand{\arraystretch}{1.5} 
\begin{table*}[htbp!]
\caption{Reconstruction RMSE obtained exploiting various approaches (DDMD, RDMD and NLPOD) using different models with different number of modes ($r$) \textcolor{rev1}{by using 200 snapshots with 10\% synthetic noise}. } \vspace{5pt}
\centering
\fontsize{6.5}{12}
\begin{tabular}{ |p{0.95cm}||p{1.3cm}|p{1.3cm}|p{1.3cm}|p{1.3cm}||p{1.3cm}|p{1.3cm}|p{1.3cm}|p{1.3cm}| }
 \hline
  & \multicolumn{3}{c}{\small$t=5$} & & \multicolumn{3}{c}{\small$t=8$} & \\ \hline
 {\scriptsize Method} & \small$r=10$ & \small$r=20$ & \small$r=50$ & \small$r=80$ & \small$r=10$ & \small$r=20$ & \small$r=50$ & \small$r=80$ \\
 \hline 
 {\scriptsize DDMD} & $3.1 \times 10^{-3}$ & $2.45\times 10^{-3}$ & $2.21\times 10^{-3}$ & $2.24\times 10^{-3}$ & $4.67\times 10^{-2}$ & $4.86\times 10^{-2}$ & $4.93\times 10^{-2}$ & $4.94\times 10^{-2}$ \\
 {\scriptsize RDMD} & $3.39\times 10^{-3}$ & $3.05\times 10^{-3}$ & $2.55\times 10^{-3}$ & $2.42\times 10^{-3}$ & $4.62\times 10^{-2}$ & $4.75\times 10^{-2}$ & $4.87\times 10^{-2}$ & $4.88\times 10^{-2}$ \\
 {\scriptsize NLPOD} & $9.52\times 10^{-4}$ & $2.76\times 10^{-4}$ & $8.34\times 10^{-4}$ & $3.44\times 10^{-3}$ & $5.02\times 10^{-2}$ & $4.9\times 10^{-2}$ & $5.1\times 10^{-2}$ & $5.41\times 10^{-2}$ \\
 \hline
\end{tabular}
\label{table:comparison10x}
\end{table*}

\begin{figure}[H]
\includegraphics[scale=0.4]{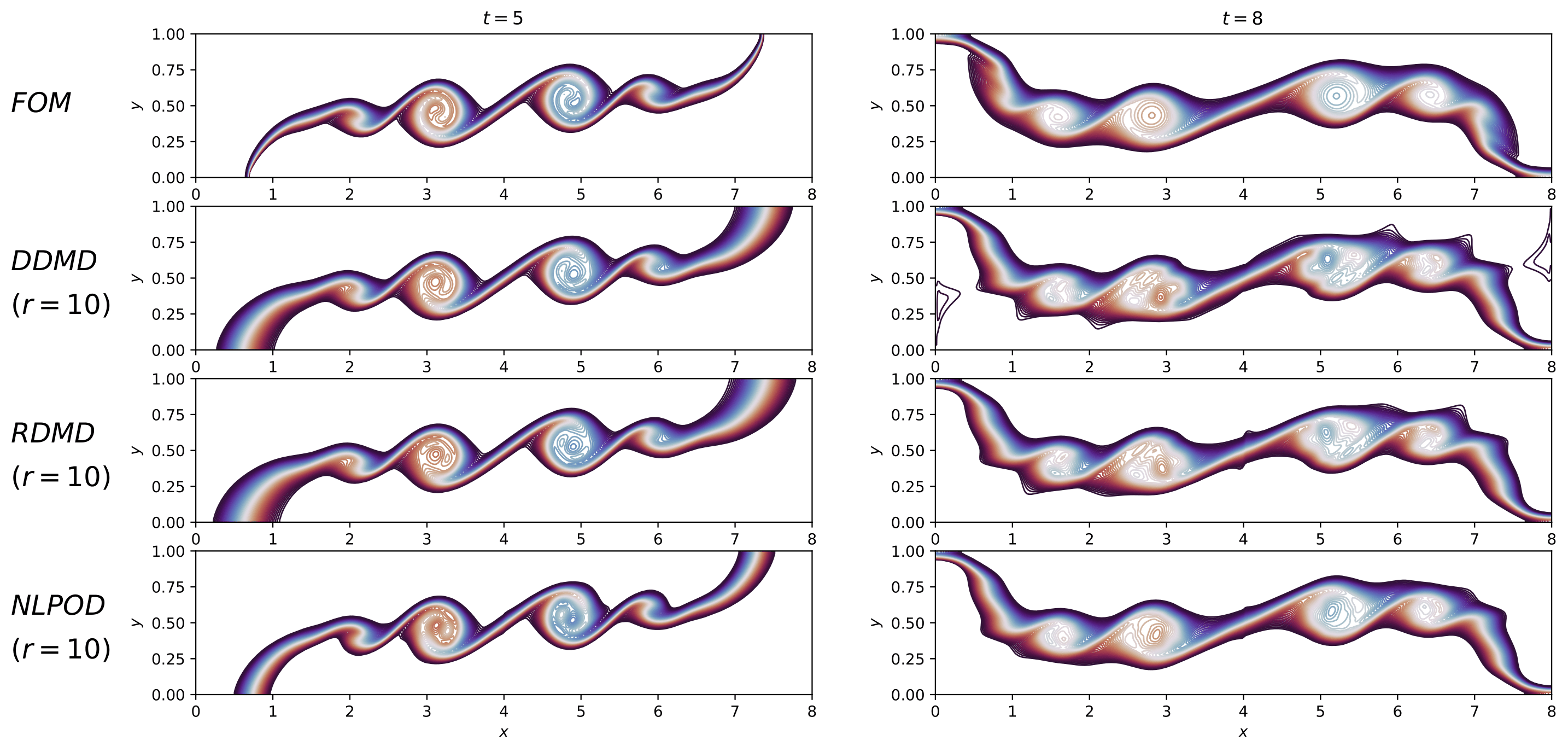}
\caption{Temperature field in snapshots $t=5$ and $t=8$ reconstructed by different methods for $r=10$ without noise.}
\label{fig:Fig1}
\end{figure}
\vspace{10pt}

\begin{figure}[H]
\includegraphics[scale=0.4]{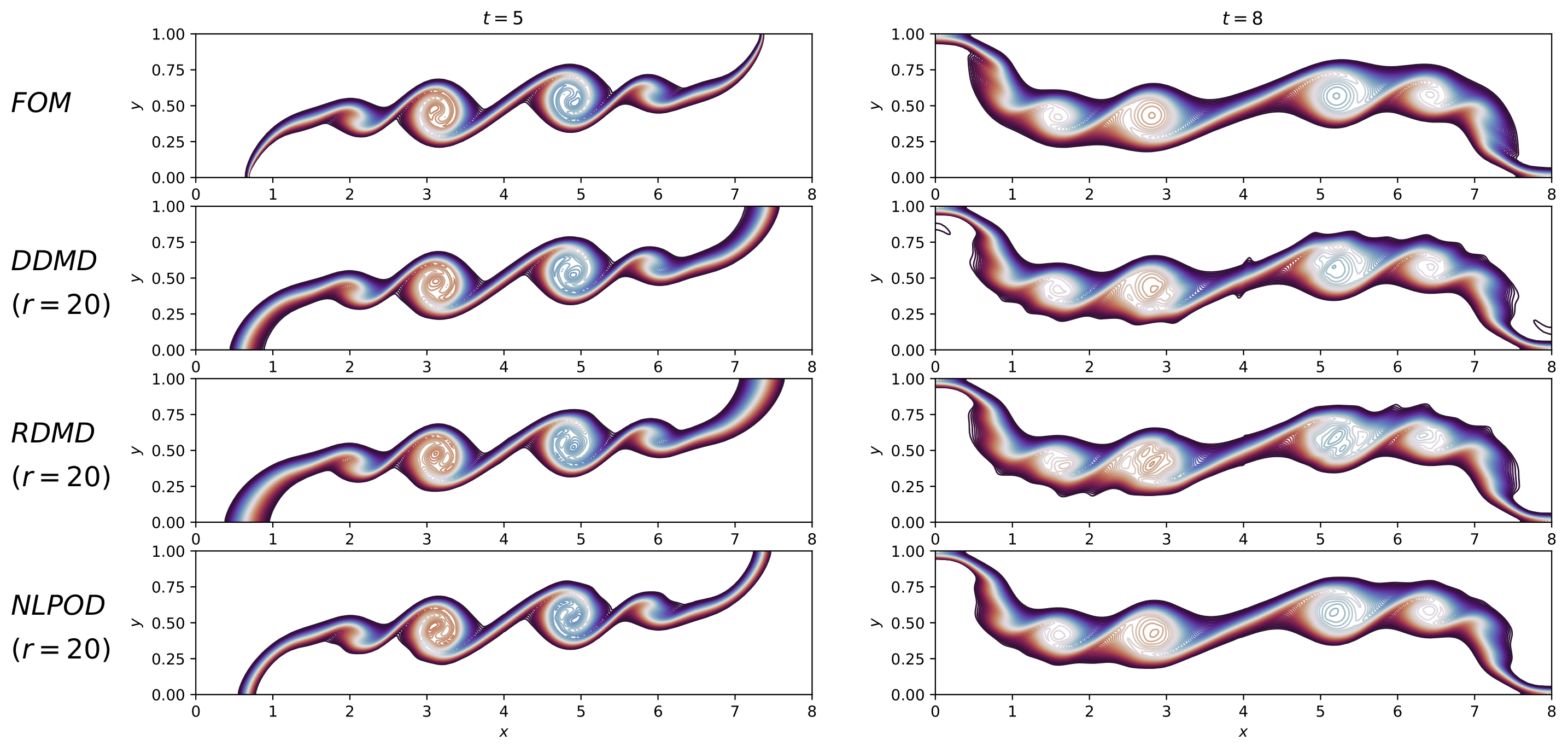}
\caption{Temperature field in snapshots $t=5$ and $t=8$ reconstructed by different methods for $r=20$ without noise.}
\label{fig:Fig2}
\end{figure}
\vspace{10pt}

\begin{figure}[H]
\includegraphics[scale=1]{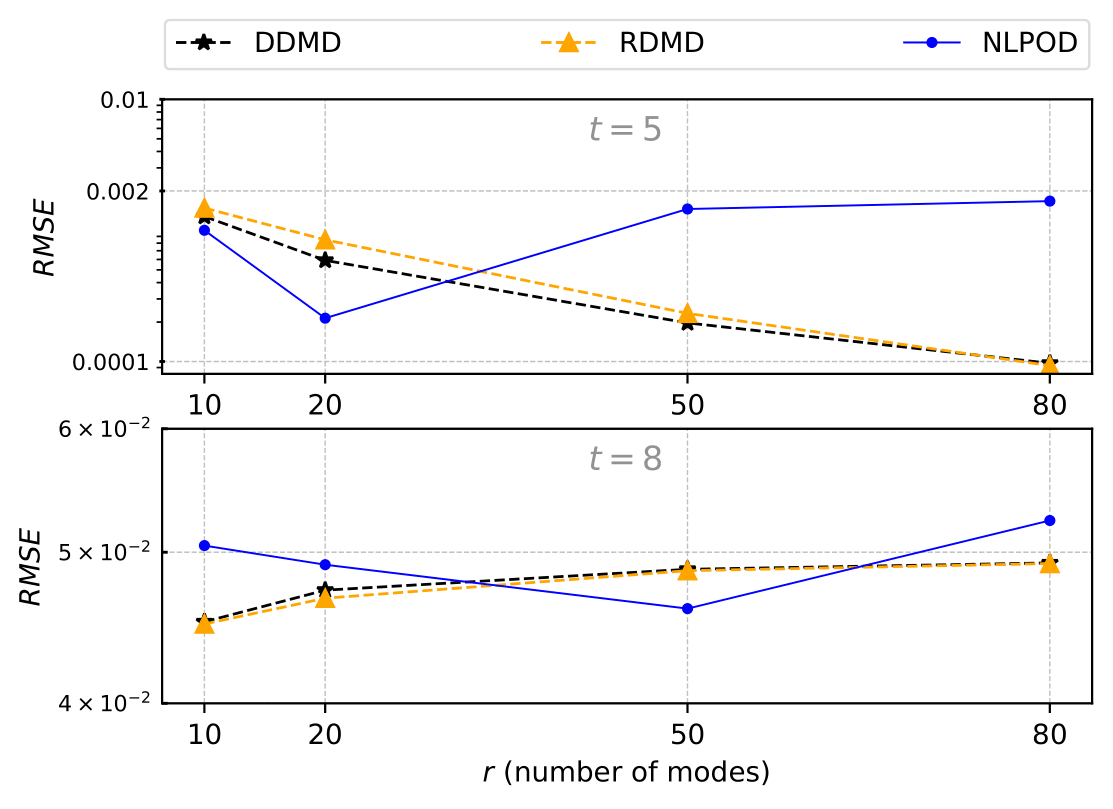}
\caption{RMSE of models reconstructed using different number of modes in two different times, $t=5$ (top) and $t=8$ (bottom) in logarithmic scale (without noise).}
\label{fig:Fig3}
\end{figure}
\vspace{10pt}

\begin{figure}[H]
\includegraphics[scale=0.4]{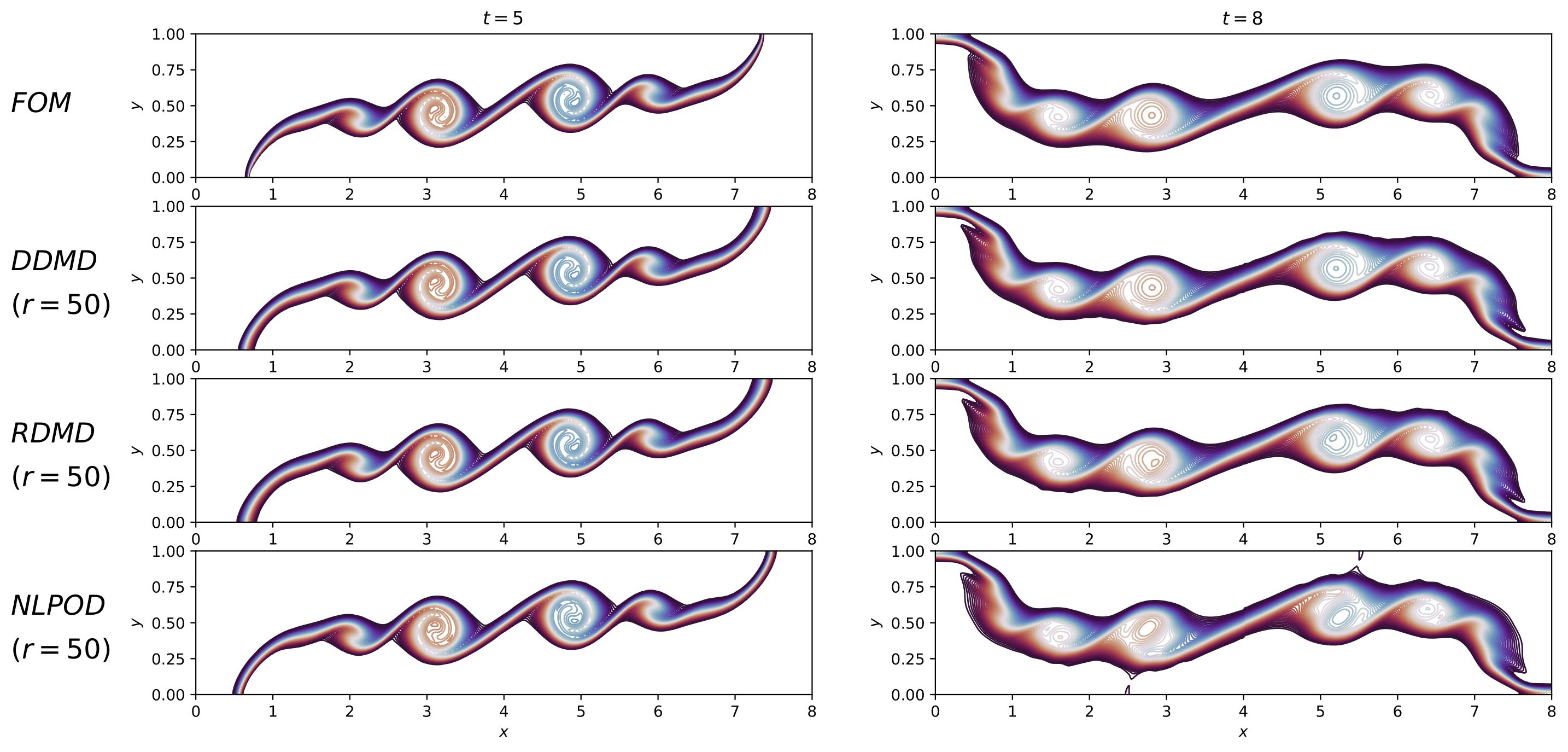}
\caption{Temperature field in snapshots $t=5$ and $t=8$ reconstructed by different methods for $r=50$ without noise.}
\label{fig:Fig4}
\end{figure}
\vspace{10pt}

\begin{figure}[H]
\includegraphics[scale=0.4]{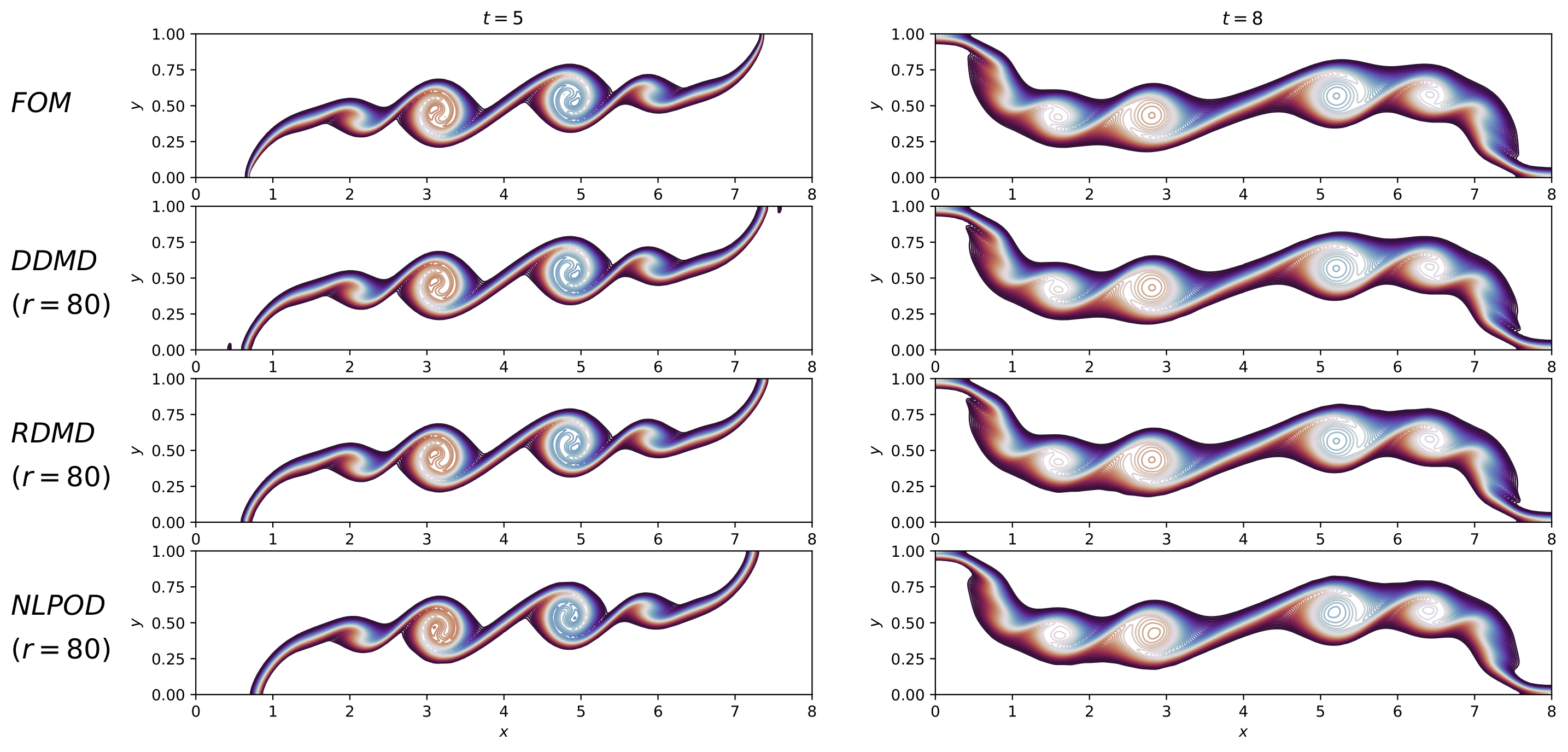}
\caption{Temperature field in snapshots $t=5$ and $t=8$ reconstructed by different methods for $r=80$ without noise.}
\label{fig:Fig5}
\end{figure}
\vspace{10pt}

\begin{figure}[H]
\includegraphics[scale=0.4]{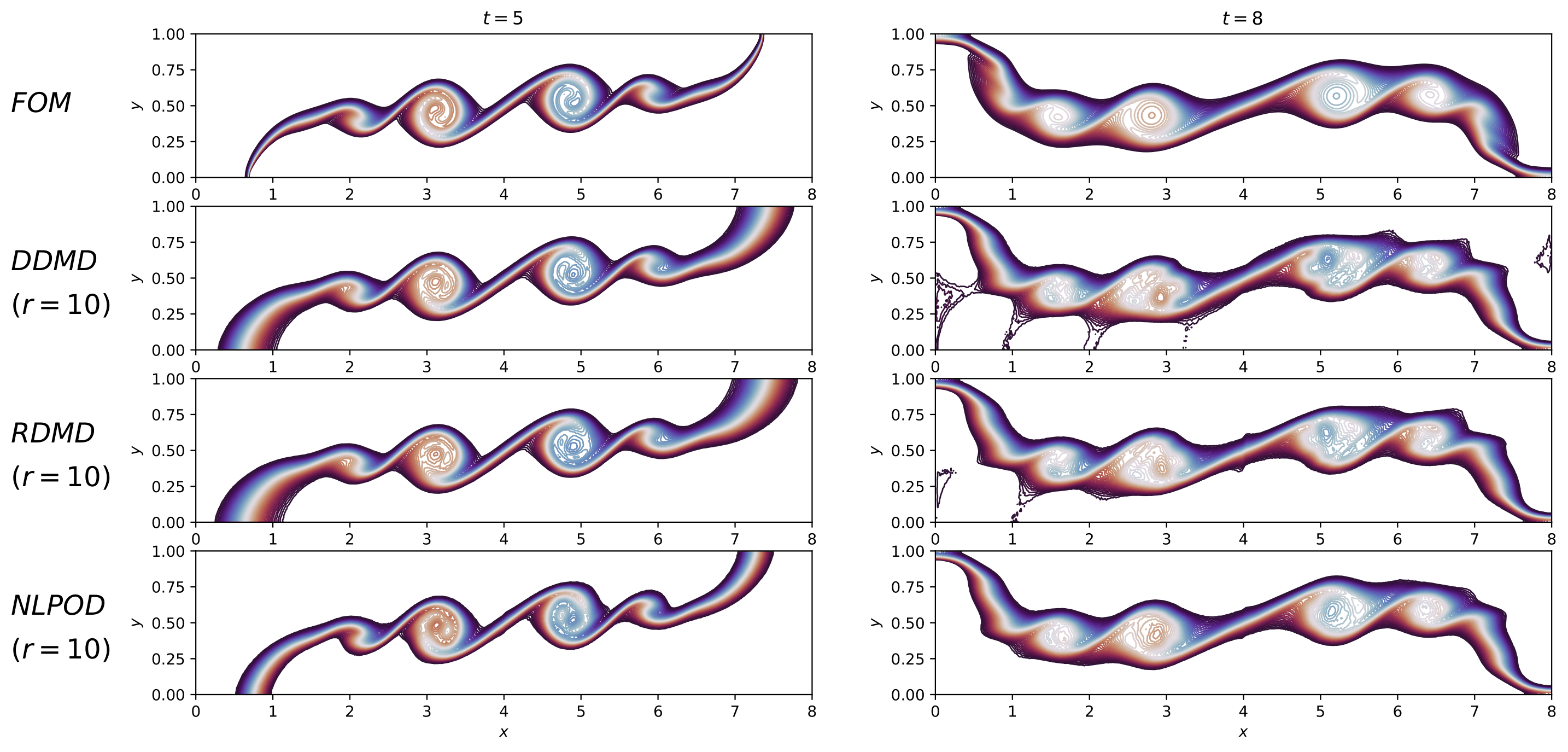}
\caption{Temperature field in snapshots $t=5$ and $t=8$ reconstructed by different methods for $r=10$ with 2\% synthetic noise.}
\label{fig:Fig6}
\end{figure}
\vspace{10pt}

\begin{figure}[H]
\includegraphics[scale=0.4]{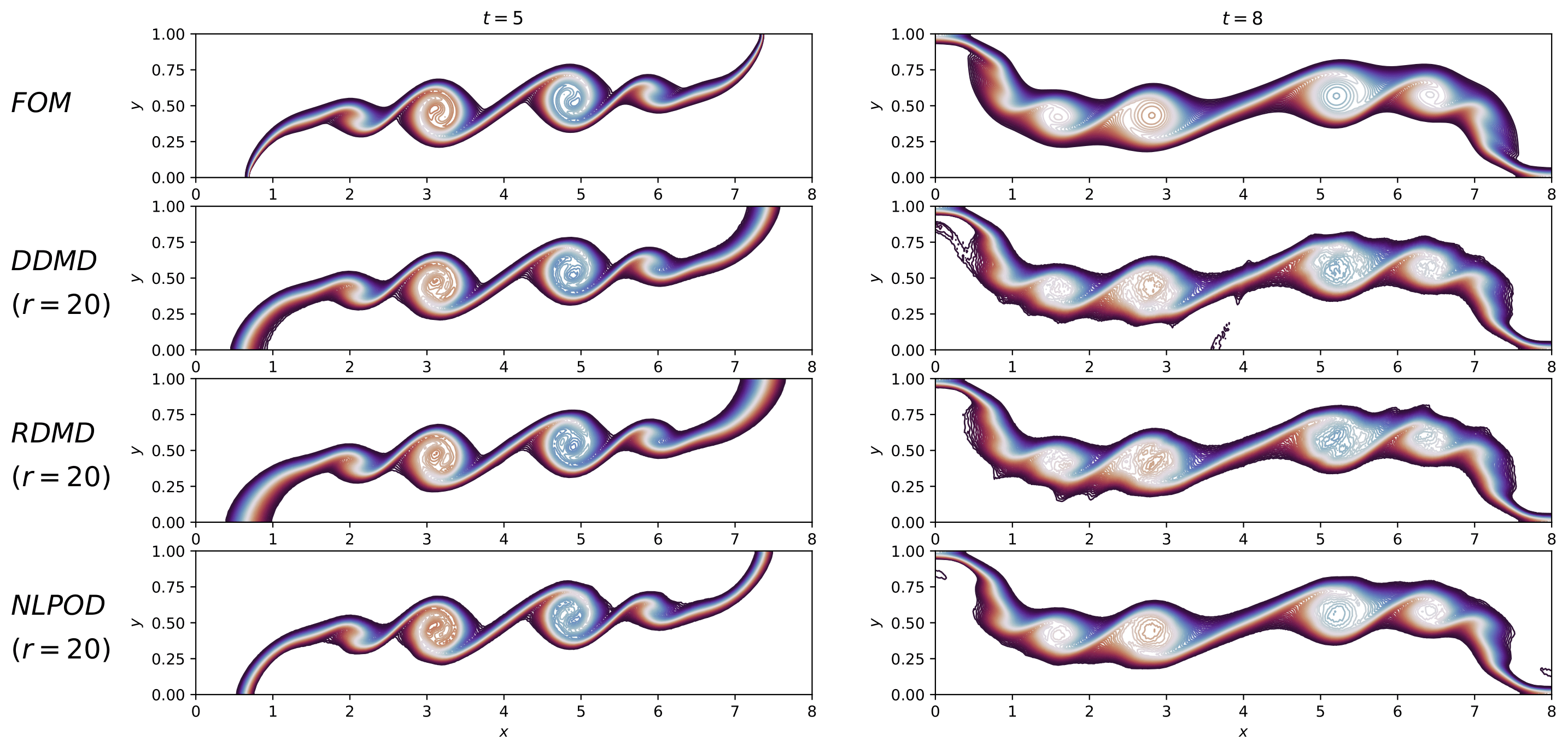}
\caption{Temperature field in snapshots $t=5$ and $t=8$ reconstructed by different methods for $r=20$ with 2\% synthetic noise.}
\label{fig:Fig7}
\end{figure}
\vspace{10pt}

\begin{figure}[H]
\includegraphics[scale=0.4]{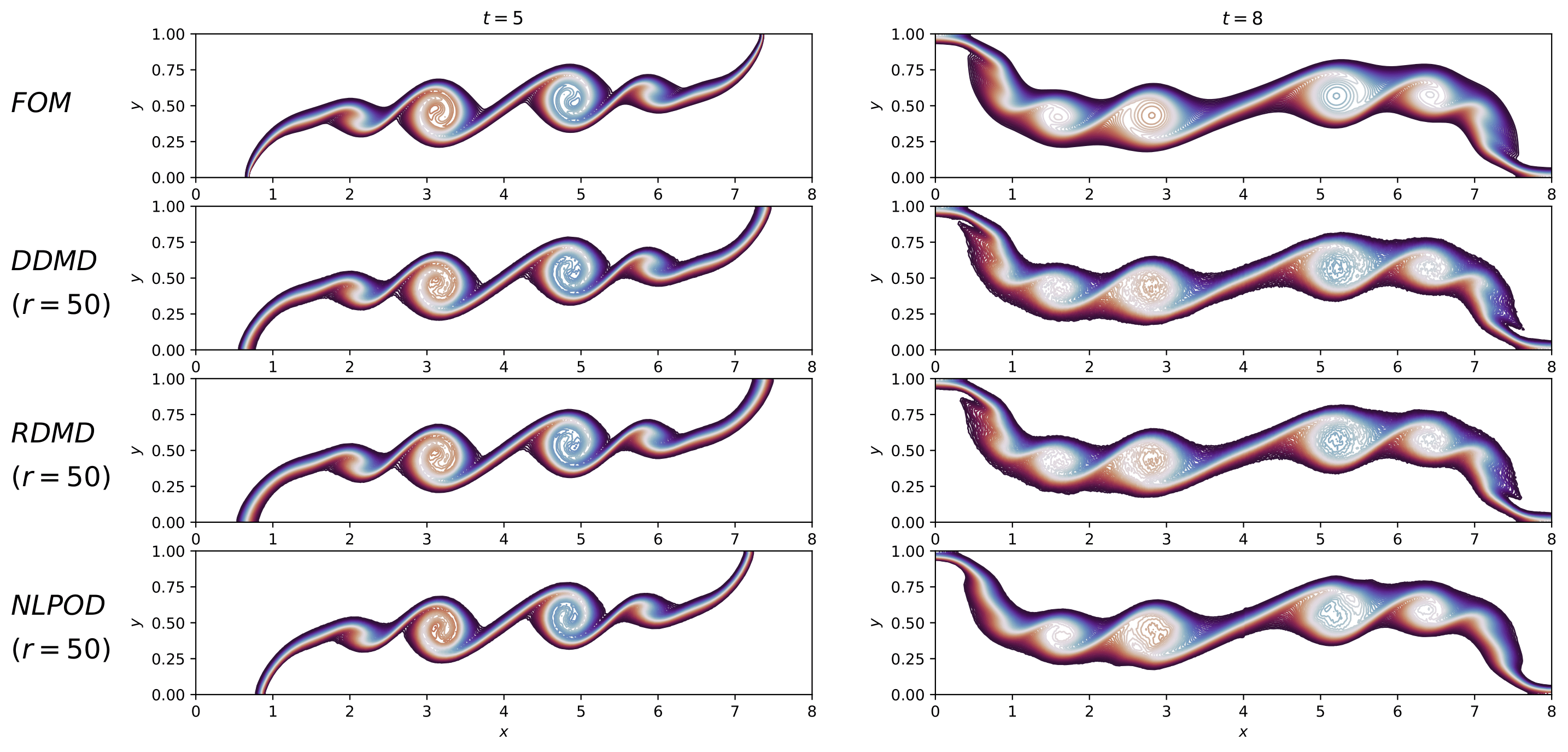}
\caption{Temperature field in snapshots $t=5$ and $t=8$ reconstructed by different methods for $r=50$ with 2\% synthetic noise.}
\label{fig:Fig8}
\end{figure}
\vspace{10pt}

\begin{figure}[H]
\includegraphics[scale=0.4]{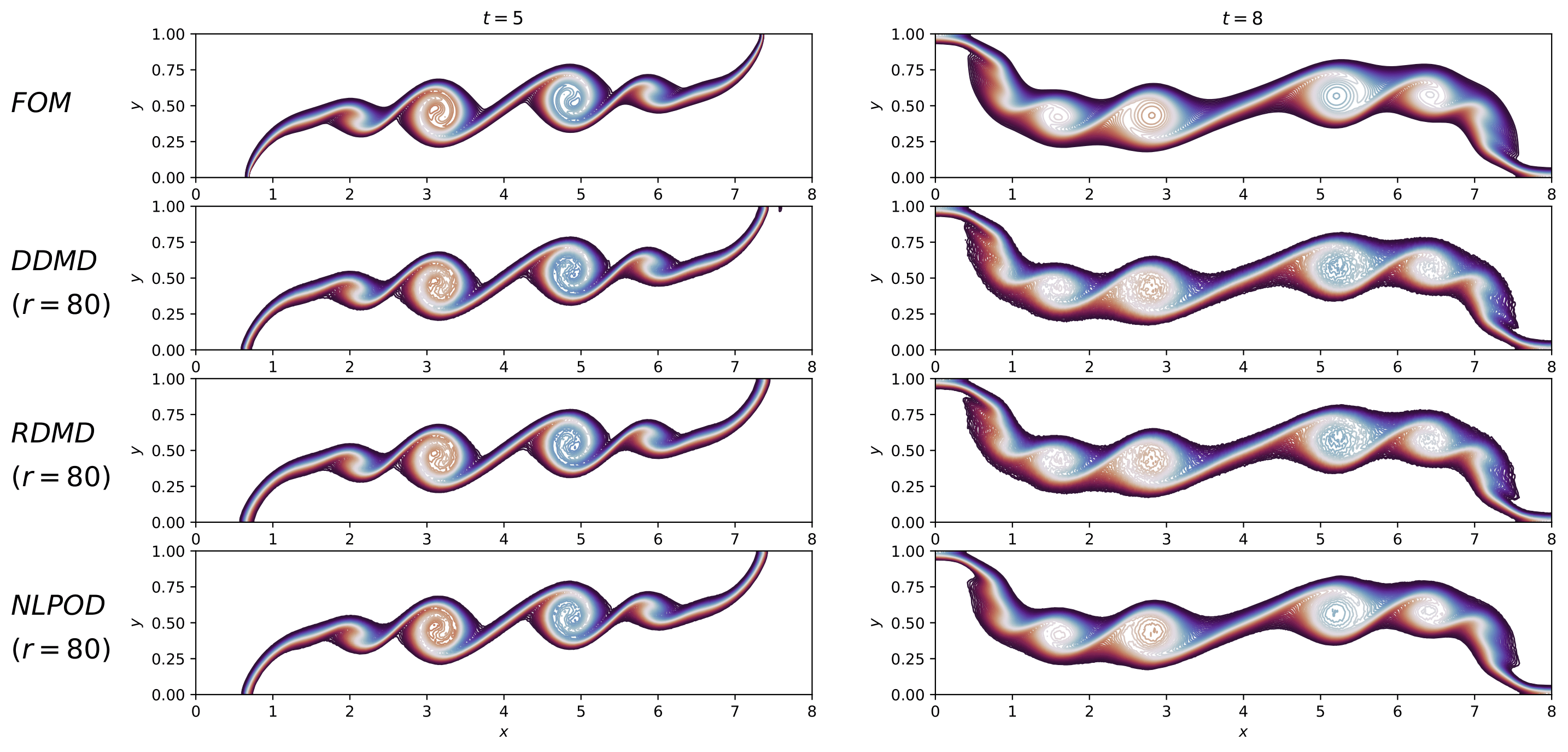}
\caption{Temperature field in snapshots $t=5$ and $t=8$ reconstructed by different methods for $r=80$ with 2\% synthetic noise.}
\label{fig:Fig9}
\end{figure}
\vspace{10pt}

\begin{figure}[H]
\includegraphics[scale=1]{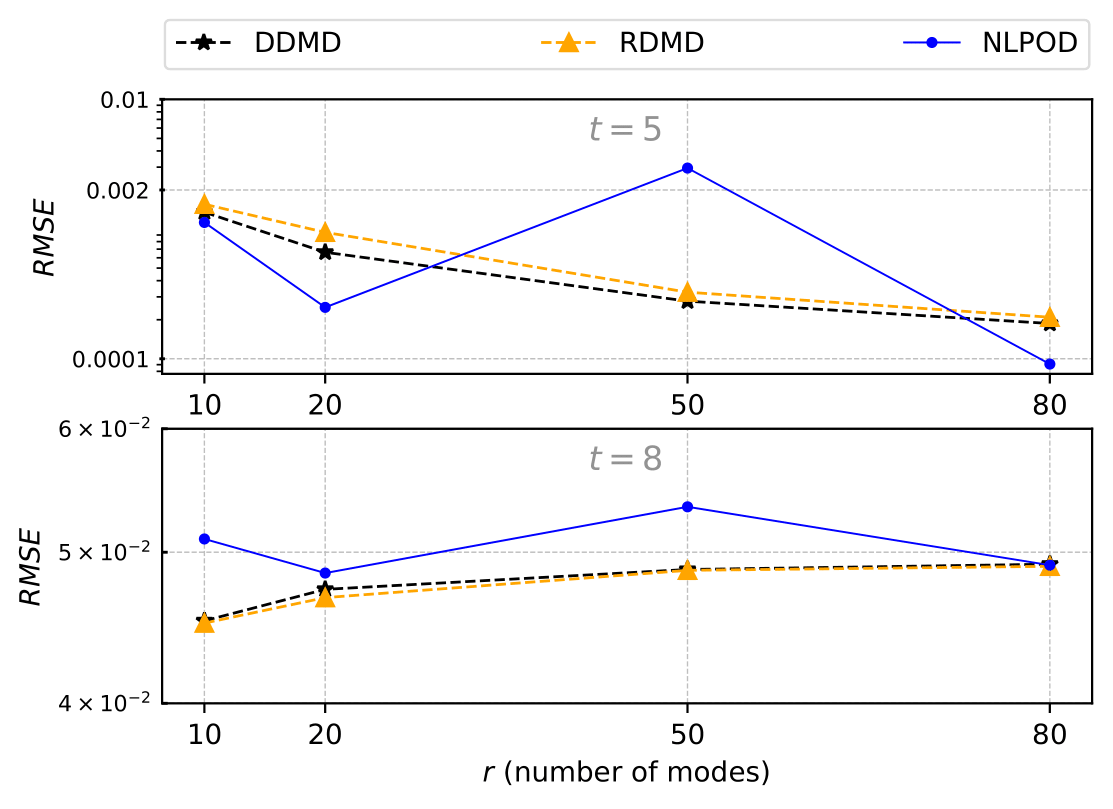}
\caption{RMSE of models reconstructed using different number of modes in two different times, $t=5$ (top) and $t=8$ (bottom) in logarithmic scale (with 2\% synthetic noise).}
\label{fig:Fig10}
\end{figure}
\vspace{10pt}

\begin{figure}[H]
\includegraphics[scale=0.4]{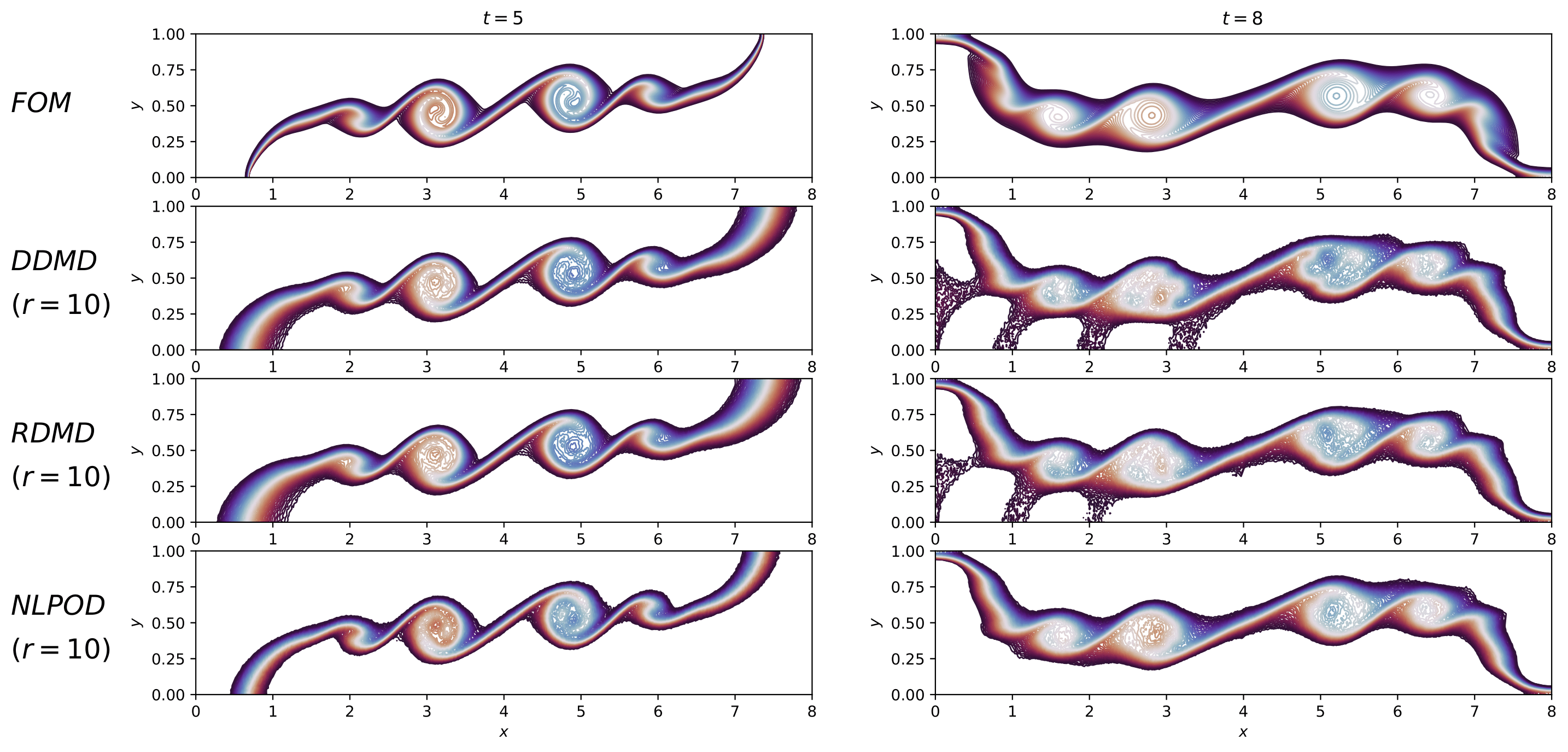}
\caption{Temperature field in snapshots $t=5$ and $t=8$ reconstructed by different methods for $r=10$ with 5\% synthetic noise.}
\label{fig:Fig11}
\end{figure}
\vspace{10pt}

\begin{figure}[H]
\includegraphics[scale=0.4]{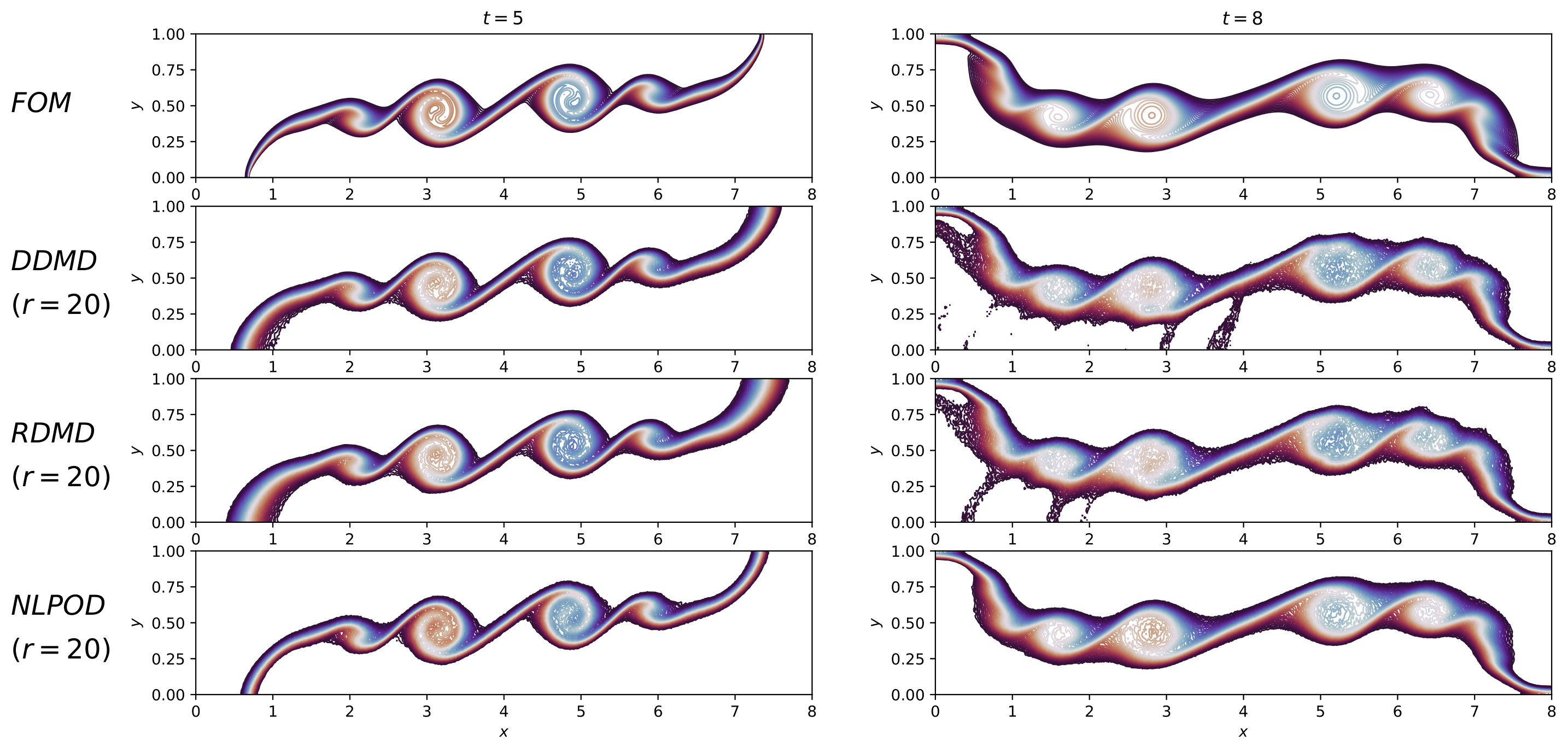}
\caption{Temperature field in snapshots $t=5$ and $t=8$ reconstructed by different methods for $r=20$ with 5\% synthetic noise.}
\label{fig:Fig12}
\end{figure}
\vspace{10pt}

\begin{figure}[H]
\includegraphics[scale=0.4]{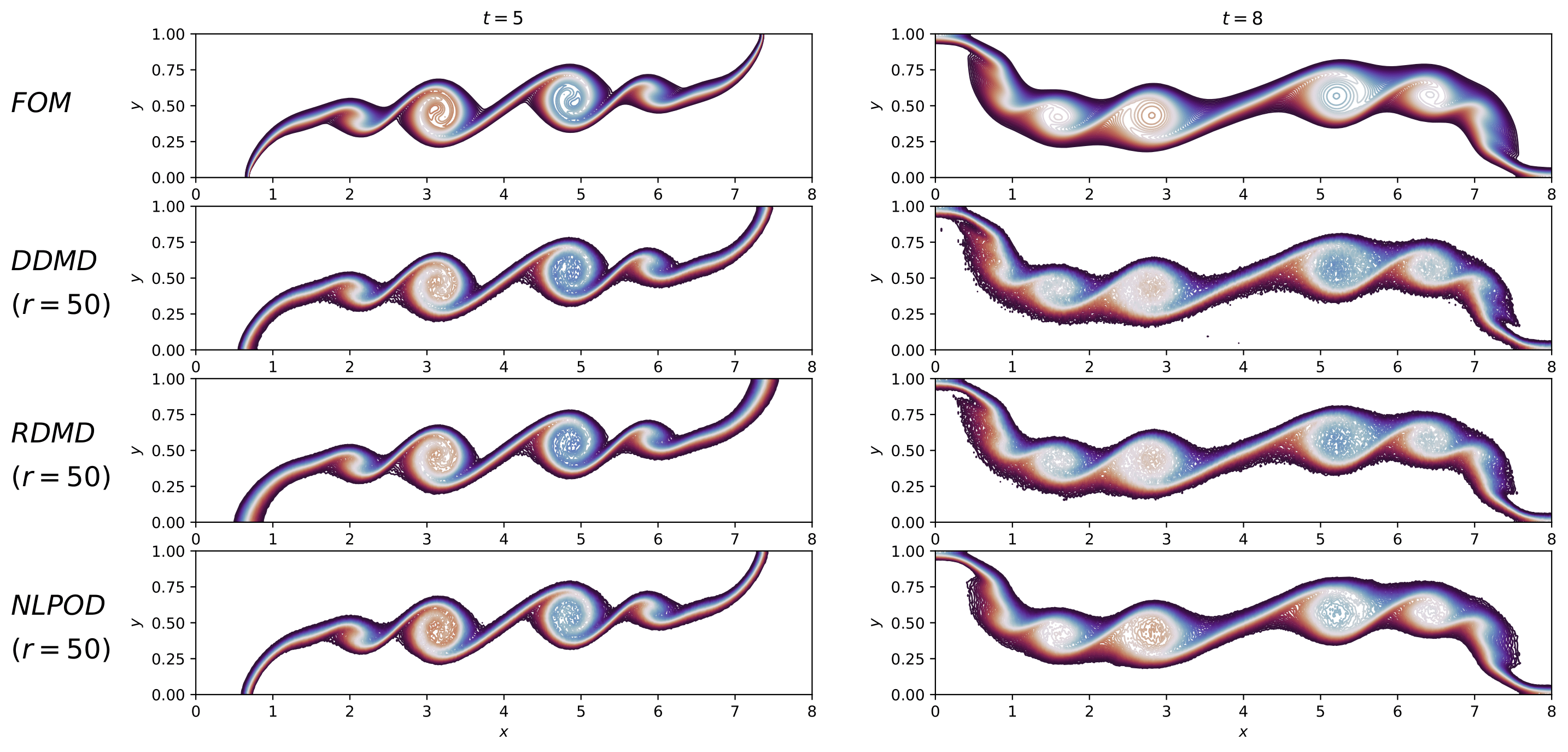}
\caption{Temperature field in snapshots $t=5$ and $t=8$ reconstructed by different methods for $r=50$ with 5\% synthetic noise.}
\label{fig:Fig13}
\end{figure}
\vspace{10pt}

\begin{figure}[H]
\includegraphics[scale=0.4]{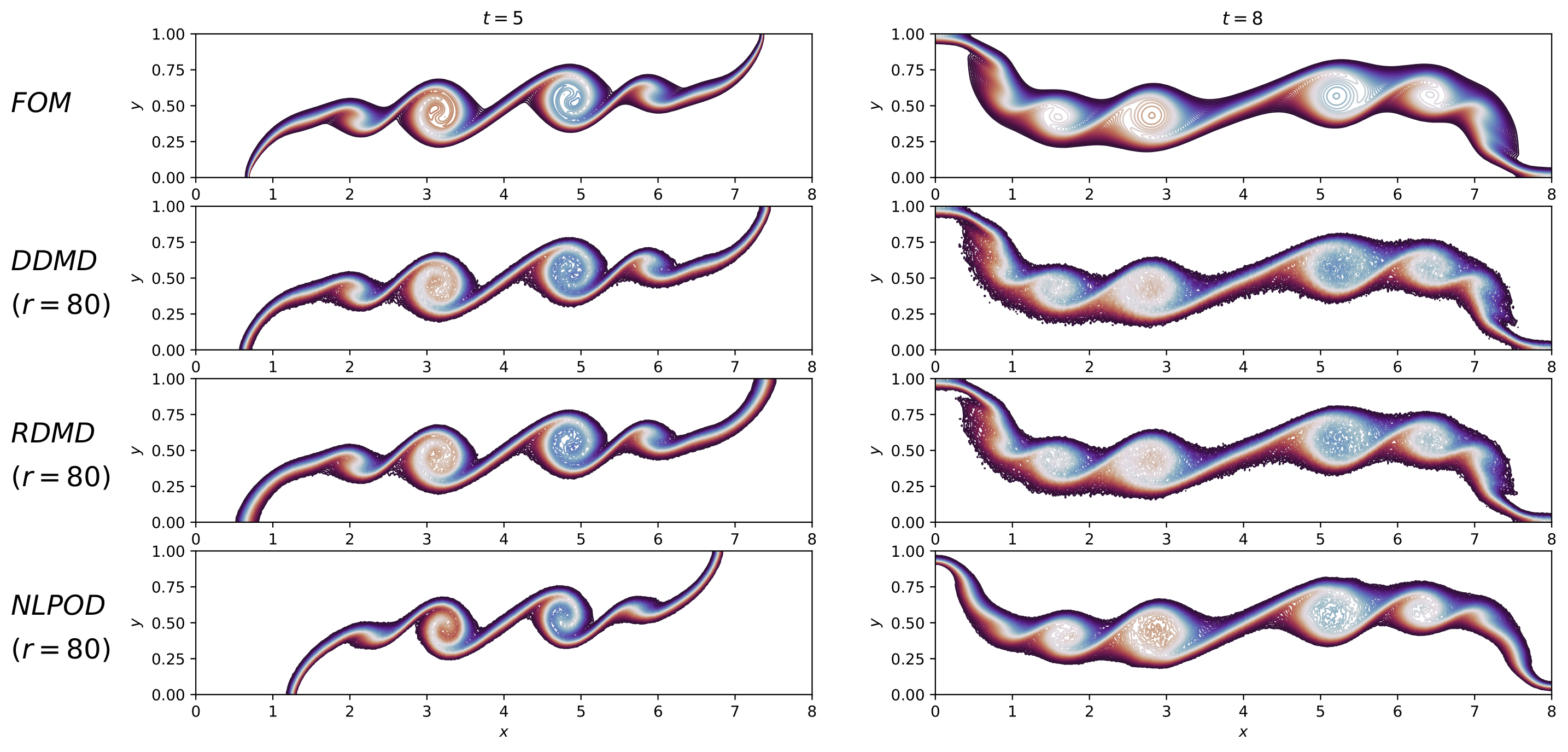}
\caption{Temperature field in snapshots $t=5$ and $t=8$ reconstructed by different methods for $r=80$ with 5\% synthetic noise.}
\label{fig:Fig14}
\end{figure}
\vspace{10pt}

\begin{figure}[H]
\includegraphics[scale=1]{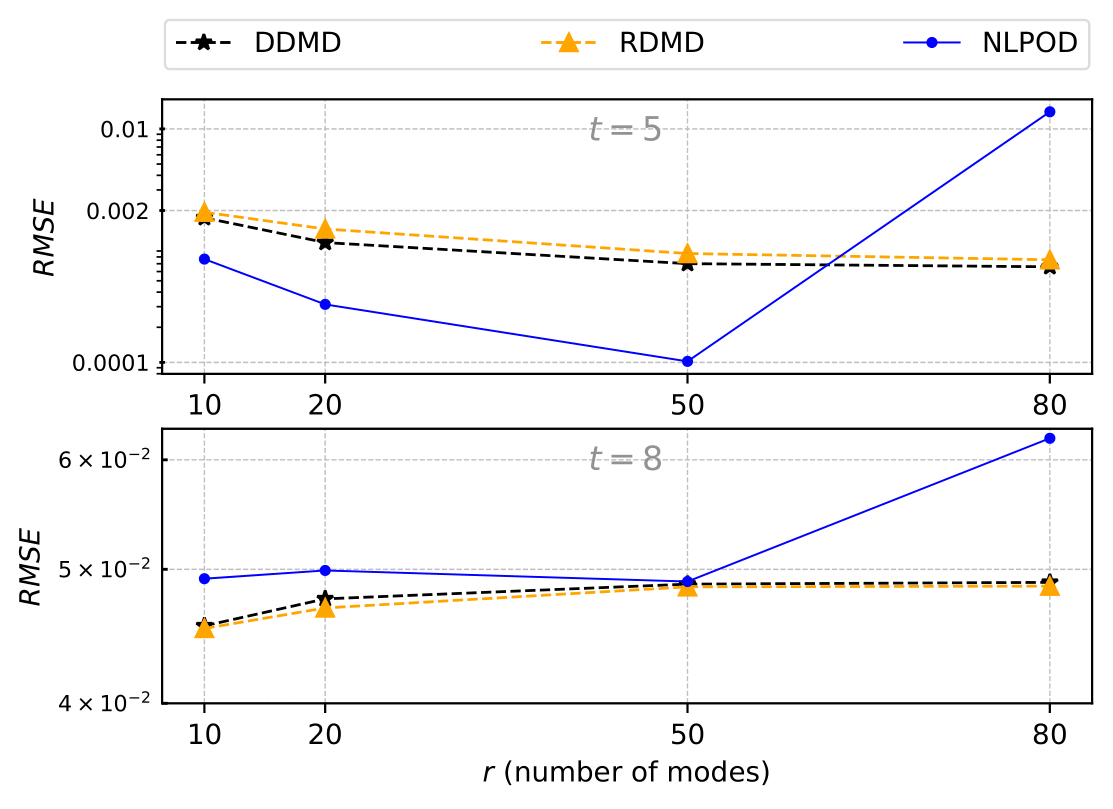}
\caption{RMSE of models reconstructed using different number of modes in two different times, $t=5$ (top) and $t=8$ (bottom) in logarithmic scale (with 5\% synthetic noise).}
\label{fig:Fig15}
\end{figure}
\vspace{10pt}

\begin{figure}[H]
\includegraphics[scale=0.4]{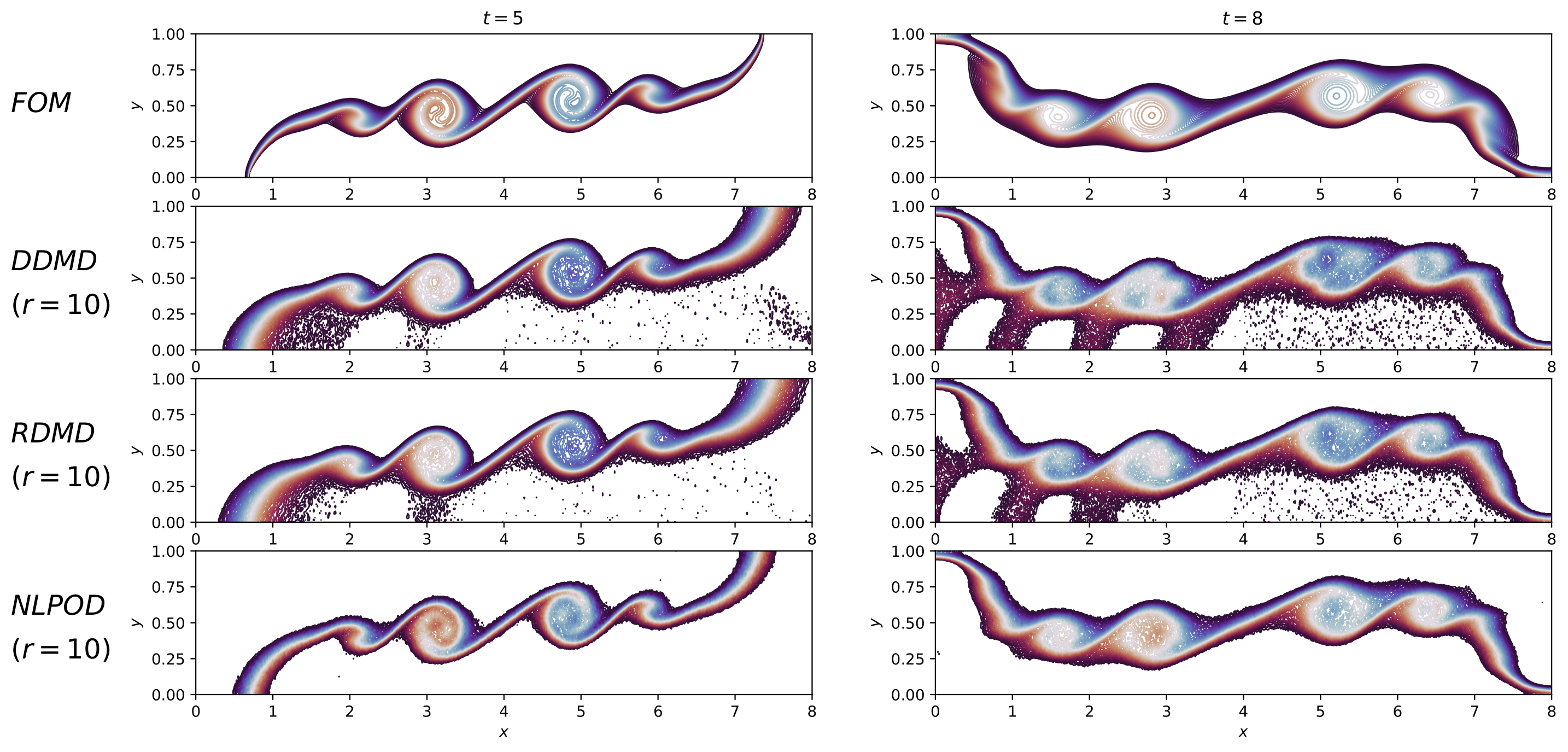}
\caption{Temperature field in snapshots $t=5$ and $t=8$ reconstructed by different methods for $r=10$ with 10\% synthetic noise.}
\label{fig:Fig16}
\end{figure}
\vspace{10pt}

\begin{figure}[H]
\includegraphics[scale=0.4]{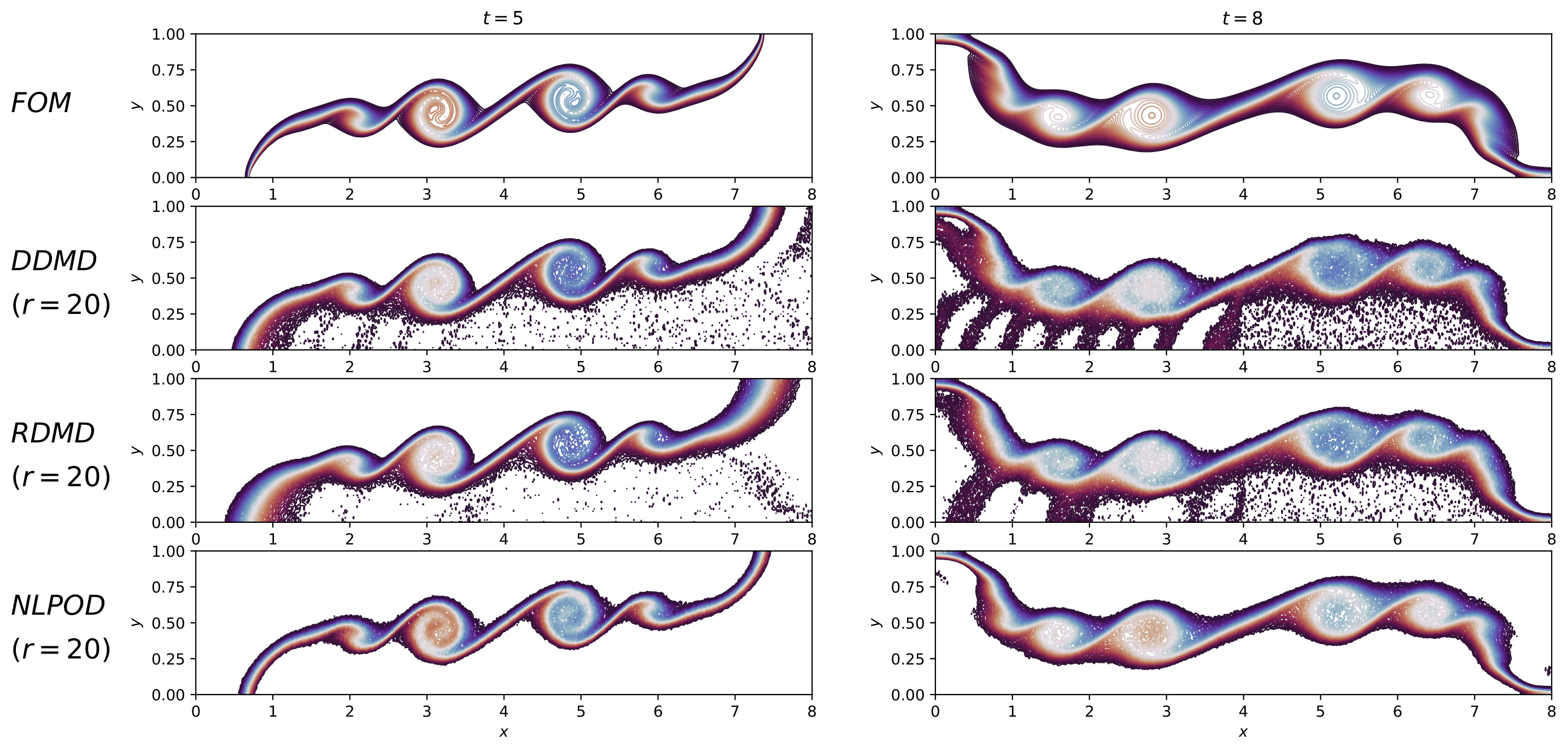}
\caption{Temperature field in snapshots $t=5$ and $t=8$ reconstructed by different methods for $r=20$ with 10\% synthetic noise.}
\label{fig:Fig17}
\end{figure}
\vspace{10pt}

\begin{figure}[H]
\includegraphics[scale=0.4]{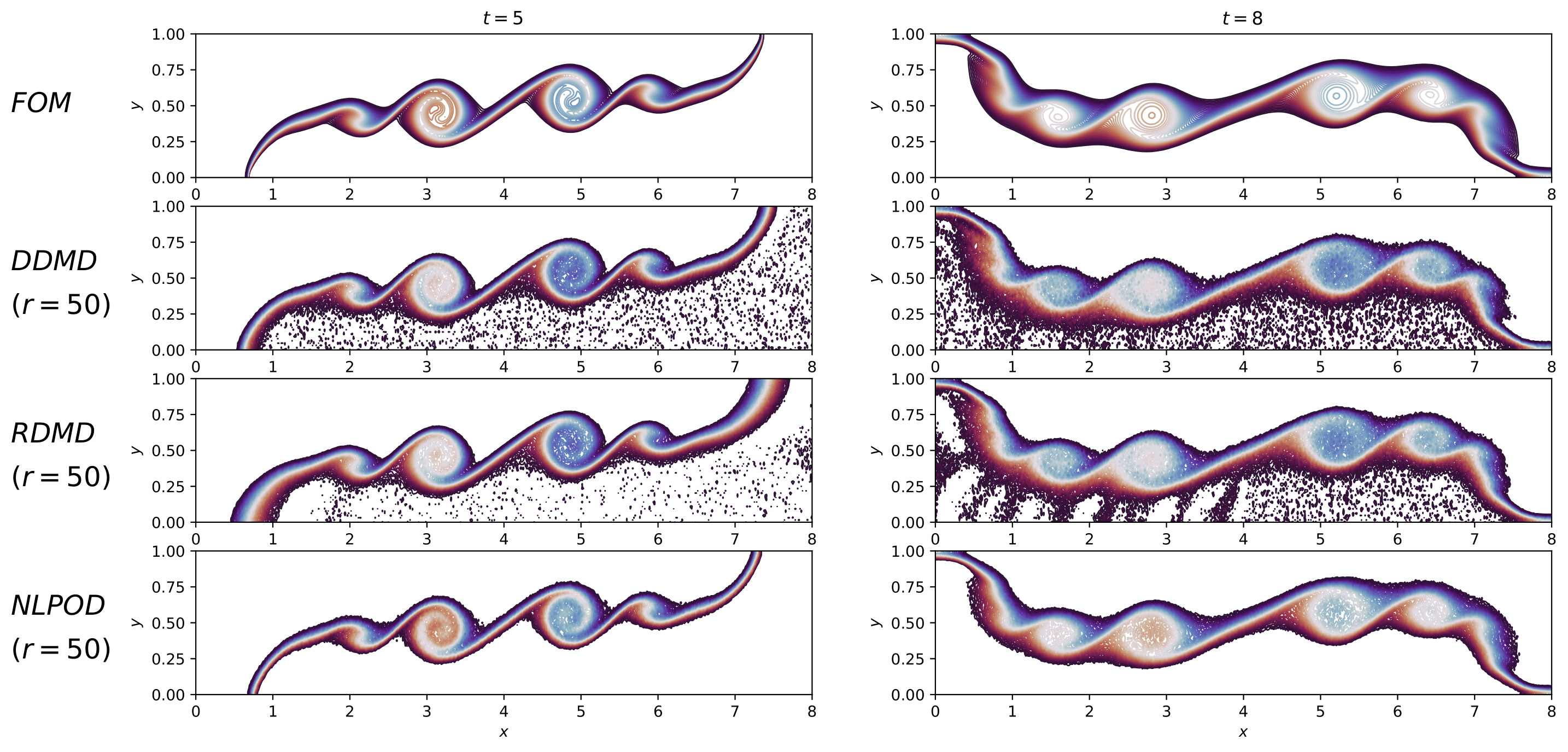}
\caption{Temperature field in snapshots $t=5$ and $t=8$ reconstructed by different methods for $r=50$ with 10\% synthetic noise.}
\label{fig:Fig18}
\end{figure}
\vspace{10pt}

\begin{figure}[H]
\includegraphics[scale=0.4]{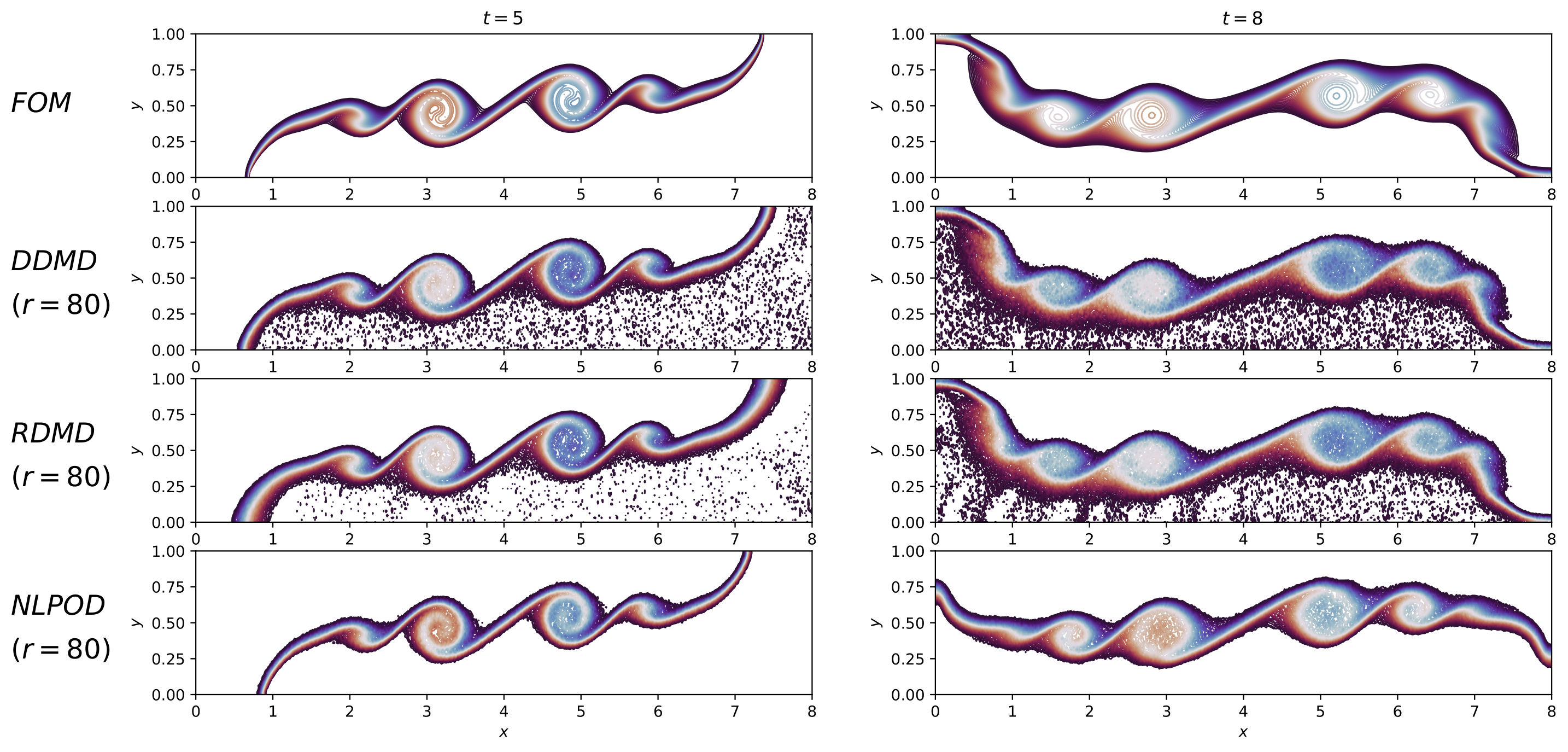}
\caption{Temperature field in snapshots $t=5$ and $t=8$ reconstructed by different methods for $r=80$ with 10\% synthetic noise.}
\label{fig:Fig19}
\end{figure}
\vspace{10pt}

\begin{figure}[H]
\includegraphics[scale=1]{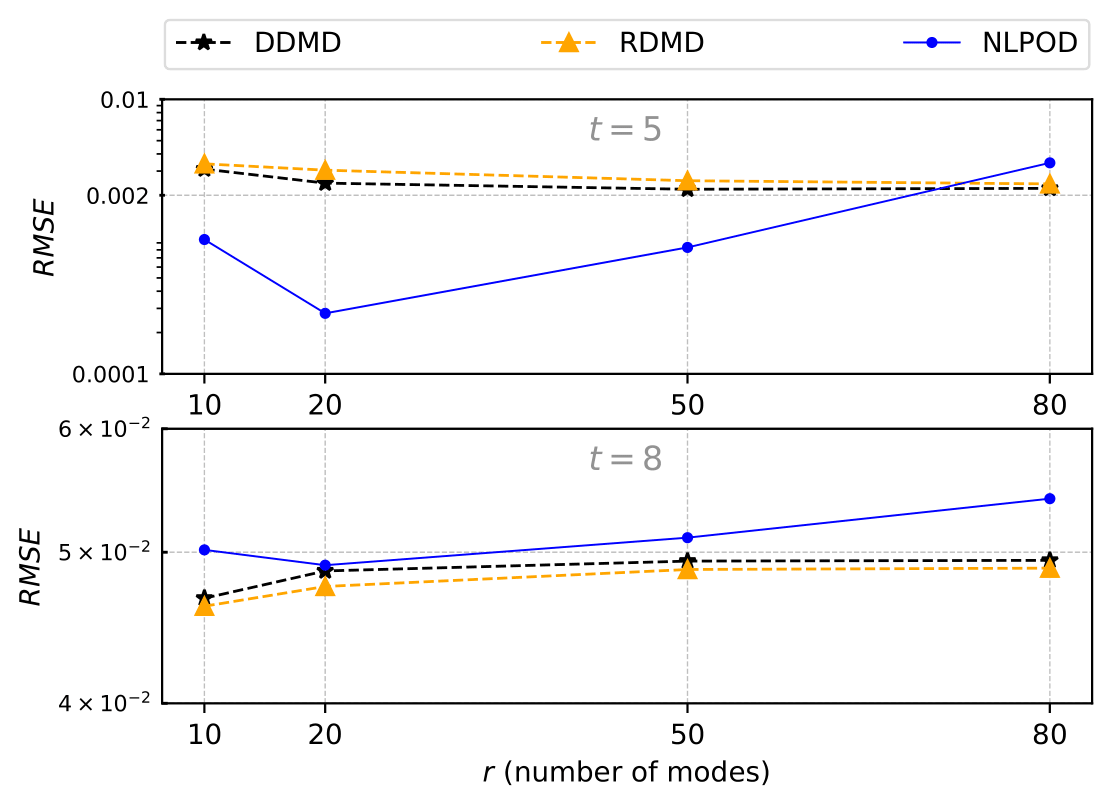}
\caption{RMSE of models reconstructed using different number of modes in two different times, $t=5$ (top) and $t=8$ (bottom) in logarithmic scale (with 10\% synthetic noise).}
\label{fig:Fig20}
\end{figure}
\vspace{10pt}

\section{Conclusions} \label{sec:con}
In this work, we compare the reconstruction capability of nonlinear proper orthogonal decomposition (NLPOD) as a new machine learning estimation approach benefiting from proper orthogonal decomposition (POD), autoencoder (AE) and long short-term memory (LSTM) neural network tools, with deterministic and randomized dynamic mode decomposition algorithms (DDMD and RDMD), respectively as two different frequency oriented ROM approaches. 
\textcolor{rev1}{We apply these methods to a convection dominated fluid flow problem governed by the Boussinesq equations where two fluid streams with two different temperatures blend together in generating vast variety of vortex patterns evolving in time.} 
The results emphasize that all approaches improve their estimation by providing them with more number of spatial modes. 
Of particular interest, we analyze the reconstruction results primarily at two different times. At the earlier time of the flow evolution where there is a relatively fewer fine scale vortical structures, our results indicate that both DMD approaches monotonically decrease errors as we increase the number of retained modes. At a later stage, however, DMD reconstruction error increases with increasing the number of modes. In contrast, we do not observe such behavior in the NLPOD approach. These observations can be related to the suboptimal selection of the relevant hyperparamaters that intensify overfitting issues. 

We also found that the NLPOD approach yields more accurate results with fewer number of modes. Increasing the number of modes may require a deeper neural network architecture for NLPOD. 
Instead of a detailed analysis on NLPOD hyperparameters, in this work, we rather focus on comparing NLPOD and DMD nonintrusive modeling approaches. It is worth to note that the difference between NLPOD and the DMD methods is basically originated from spatial modes and temporal coefficients. Despite spatial modes are obtained from different procedures within the NLPOD and DMD procedures, their structure is similar as they are created \emph{first} by applying the SVD onto the snapshot data set to generate a nonintrusive ROM. Thus, the main difference relies on the formulation of the dynamics. In NLPOD, the calculated temporal weights were found using LSTM that is trained to predict the time evolution of latent space variables representing a compressed version of the POD coefficients. On the other hand, the DMD temporal behavior is estimated by the products of the eigenvalue decomposition of the approximated Koopman operator.
Besides, the results of temperature field reconstruction show that the NLPOD is more efficacious in reconstructing the boundaries compared to the DMD. Also, it can be seen that the NLPOD estimates smoother edges than the DMD with the same number of modes. This superiority can be attributed to the improved representation potential of the NLPOD in estimating the weights since the NLPOD, unlike the DMD, exploits a nonlinear approach to find the weights of modes.

\textcolor{rev1}{To evaluate the noise sensitivity of the models, they are tested by feeding them with noisy input. We found that the NLPOD can reconstruct models more effectively than DMD and that it can be a reliable model if the cancellation or attenuation of the noise is an issue. In addition, adding noise to both DMD methods results in errors with smaller variances as the number of modes changes. In fact, in the presence of noise, increasing the number of modes does not have a significant effect on reducing the error.}



\section*{Data availability}
The data that supports the findings of this study
is available within the article. The computer scripts and datasets used and/or analysed during the current study are available from the corresponding author on reasonable request. 

\section*{Disclosure statement}

We declare we have no competing interests.

\bibliographystyle{apacite}
\bibliography{references}

\end{document}